\documentclass[12pt]{iopart}

\usepackage{graphicx}
\usepackage{caption}

\begin{document}

\title{Light weakly interacting massive particles}

\author{Graciela B. Gelmini}

\address{Department of Physics and Astronomy, University of California, Los Angeles (UCLA), 475 Portola Plaza, Los Angeles, CA 90095, USA}
\ead{gelmini@physics.ucla.edu}
\begin{abstract}
  Light WIMPs are  dark matter particle candidates with weak scale interaction with the known particles, and mass in the GeV to 10's of GeV range. Hints of light WIMPs have appeared in several dark matter searches in the last decade. The unprecedented possible coincidence into tantalizingly close regions of mass and cross section of four separate direct detection experimental hints and a potential indirect detection signal in gamma rays from the galactic center,  aroused considerable interest in our field. Even if these hints did not so far result in a discovery, they have had a significant impact in our field. Here we review the evidence for and against light WIMPs as dark matter candidates and discuss future relevant experiments and observations~\footnote{Review published in Rept.\ Prog.\ Phys.\  {\bf 80}, no. 8, 082201 (2017).}
\end{abstract}

\maketitle

\section{Introduction}
 
Observations on all astrophysical  and cosmological scales at and above the scale of dwarf galaxies
indicate that dark matter (DM) constitutes about 25\% of the content of the Universe (see e.g.~\cite{pdg-dark-matter}). Most of the matter in galaxies and galaxy clusters consists of DM. More than 80\% of the mass of our own galaxy resides in a spheroidal DM halo, which extends much beyond the visible disk of the galaxy.

 The name DM is a historically motivated misnomer for what is actually ``invisible matter", i.e. matter that has not been observed to absorb or emit light of any frequency. There is no  evidence that the DM has any other interaction but gravitational (see e.g.~\cite{Gelmini:2015zpa}). The DM cannot consist of atomic matter (normal or ``baryonic" matter, consisting of protons and neutrons),  which accounts for an additional 5\% of the content of the Universe (see e.g.~\cite{pdg-cosmological-parameters}).  The remaining 70\% consists of dark energy, a component which, unlike matter, has repulsive gravitational interactions. 

The currently most accurate determination of the composition of the Universe comes from global fits of cosmological parameters to a variety of observations, combined with General Relativity (see e.g.~\cite{Gelmini:2015zpa, pdg-cosmological-parameters}). Five independent measurements of the abundance of atomic matter  show that it amounts to less than 5\% of the content of the Universe: the X-ray emission from galaxy clusters, the relative height of the odd and even peaks in the angular power spectrum of Cosmic Microwave Background (CMB) anisotropies, the abundance of light chemical elements generated in Big Bang Nucleosynthesis (BBN), Baryon Acoustic Oscillations and absorption lines of the light of quasars (the Lyman-$\alpha$ forest). Some part of the atomic matter may well contribute to a baryonic type of DM in the form of non-luminous gas or  Massive Astrophysical Compact Halo Objects (MACHOs). This baryonic MACHOs could be condensed objects such as black holes of stellar origin, neutron stars, white dwarfs, very faint stars, brown dwarfs or planets. However, MACHOs which could constitute the DM must be non-baryonic. They could be ``Primordial Black Holes" (PBH),  hypothetical black holes proposed by Carr and Hawkings in 1974~\cite{Carr:1974nx} that could be created  in a primordial phase transition, maybe during inflation (see e.g.~\cite{Kusenko:2013saa,Green:2014faa}), always  before BBN (so that even if the matter which formed the PBH could contain quarks, the constituents of protons and neutrons,  PBH are not  part of the atomic matter). 

It is worth establishing at this point the timing of some important episodes in the history of the Universe.  All measurements so far have confirmed the Big-Bang model of a hot early Universe expanding adiabatically for most of its lifetime of $t_U=$ 13.799 $\pm 0.021 \times 10^{9}$ y~\cite{Ade:2015xua}. BBN is the earliest episode in the history of the Universe from which we have data, consisting of  the relative abundance of the light elements produced then, when neutrons and protons fused together to form D, $^4$He and $^7$Li. It  took place between 3 and 20 minutes after the Bang, when the temperature of the Universe was $T\simeq$ MeV. In order for BBN and all the subsequent  history of the Universe to proceed as we know it, it is enough that the earliest and highest temperature of the radiation-dominated period in which BBN happens is just larger than  4 MeV~\cite{Hannestad:2004px}. We do not know the thermal history of the Universe before its temperature was 4 MeV and it is worth keeping in mind that most DM candidates are produced during this pre-BBN period.

The CMB was emitted  379 ky after the Bang, when the temperature of the Universe was $T\simeq 3$ eV and atoms became stable for the first time. After this moment, called ``recombination", the Universe became populated by neutral constituents (the atoms) instead of plasma, and consequently the mean free path of photons became very large on cosmological scales. Shortly before the emission of the CMB (when the Universe was about 100 ky old and $T\simeq 1$ eV) the Universe passed from being radiation-dominated to being matter-dominated by the DM. The acoustic oscillations in the pre-recombination plasma, or Baryon Acoustic Oscillations,  which produce anisotropies in the CMB photons, also leave an imprint in the atomic matter which is seen in the distribution of galaxies in space as excesses in galaxies separated at certain distances.
 
We know many properties of the DM, although  its  nature  remains one of the most important open problems in science. 

The DM has attractive gravitational interactions, it behaves as regular matter for gravitational interactions,  and it is still present in the Universe, thus it is either stable or has a lifetime much larger than the lifetime of the Universe. Because there is no  evidence that the DM has any other interaction but gravitational, it is natural to wonder if the many observational evidences for DM are instead  showing departures from the law of gravity itself. However, no proposed modification of gravity can explain all the observational evidences for DM without introducing some form of DM too (see e.g. \cite{Gelmini:2015zpa}).

 The DM is not observed to interact with light. Thus, it should either be neutral (maybe with a small electric or magnetic dipole moment~\cite{Pospelov:2000bq, Sigurdson:2004zp} or an anapole moment~\cite{Pospelov:2000bq, Ho:2012bg}) or have a small effective electric charge~\cite{Feldman:2007wj}.  A small electromagnetic coupling could arise  if the DM is part of a ``dark sector", a set of new particles not charged directly under any of the Standard Model forces (weak, strong, and electromagnetic) which couple to the known particles  only though the small admixture of a ``dark photon" with  the usual photon or the neutral mediator of weak interactions, the Z boson (see e.g.~\cite{Alexander:2016aln}). 
 
The bulk of the DM must be dissipationless, i.e. it cannot cool by radiating as atomic matter does  to collapse in the center of  disk galaxies. Otherwise,  the extended galactic dark halos would not exist.
 However,  a small fraction of the DM, 5  to 10\%, similar to the fraction of atomic matter in a galaxy,  could be dissipative without disrupting the halo~\cite{DDDM}.

The DM has been mostly assumed to be collisionless, however the upper limit on  the DM self-interaction cross section $\sigma_{\rm self}$ is  huge. An upper limit on the $\sigma_{\rm self}/m$ ratio, where $m$ is the DM constituent mass, comes from a lower limit on the DM mean free path $\lambda_{mfp} \simeq 1/ n \sigma_{\rm self} = m/ \rho \sigma_{\rm self}$ in structures of known density $\rho$~\cite{Clowe:2006eq} ($n$ is the number density $n= \rho/m$). The limit is $\sigma_{\rm self}/m\leq$ 2 barn/GeV $\simeq$ 2$ \times 10^{-24}$ cm$^2$/GeV.  By comparison, the neutron capture cross section of uranium  is a few barns!      
  DM with a large self-interaction cross section close to this limit is called ``Self-Interacting DM" (SIDM)~\cite{SIDM, Zavala:2012us}. DM with less than an order of magnitude smaller $\sigma_{\rm self}$ is indistinguishable from collissionless. 

\begin{figure}[t]
\includegraphics[width=1.02\textwidth]{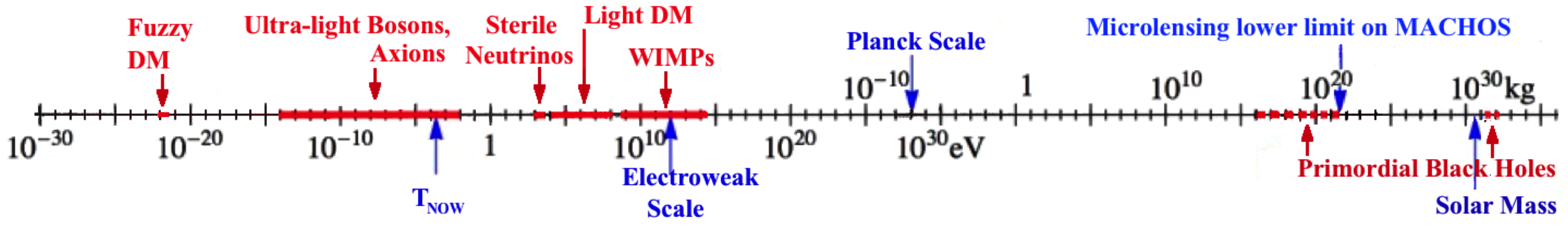}
\caption{Mass ranges (solid red lines) of several DM candidates. Starting from the lowest mass: ``Fuzzy DM"~\cite{fuzzy-DM}; ultra-light bosons, such as Axions and Axion-Like Particles (ALPs) which are scalar bosons, and also dark photons (also called hidden-sector photons) which are vector bosons  and together with ALPs  are  sometimes called Weakly Interacting Slim Particles or WISPs  
(see e.g.~\cite{Ringwald:2012hr} and Sect.~6 of~\cite{Hewett:2012ns}); sterile neutrinos (see e.g.~\cite{Gelmini:2015zpa}); Light DM (LDM), also called Sub-GeV WIMPs 
(see e.g. Sect. V of~\cite{Alexander:2016aln}); WIMPs (see e.g.~\cite{Gelmini:2015zpa}); the potential windows for Primordial Black Holes (dashed red lines) would require evading or weakening some existing bounds to account for the whole of the DM within them (see e.g. Section VI of~\cite{Carr:2016drx}).}
\label{Fig-0}
\end{figure}

This brings us to what we know about the mass of the DM constituents. An upper limit on the mass of the major constituent of  the DM $ m \leq 2 \times  10^{-9}$ M$_\odot= 2 \times 10^{48}$ GeV  at the 95\%C.L.
(M$_\odot$ is a solar mass) comes from unsuccessful searches for MACHOS  in the dark halo of our galaxy using gravitational microlensing  (by the ground-based MACHO, EROS and OGLE surveys~\cite{Alcock:1998fx} and the Kepler satellite~\cite{Griest:2014cqa}), combined with bounds on the granularity of the DM  (starting with wide binary stars disruption limits for $m> 30$ M$_\odot$~\cite{granularity}).
Microlensing is a type of gravitational lensing in which the multiple images of the lensed star are superposed, producing a temporary magnification of the star flux if an object passes near the line of sight to the star.  Above this limit, MACHOS  can account for only a small fraction of the dark halo of our galaxy, except perhaps at the cross-over from the microlensing and wide binaries disruption limits. Weakening somewhat both these limits, a window from 20 M$_\odot$ to 100 M$_\odot$ has been suggested to exist, in which MACHOS could constitute all of the DM~\cite{Bird:2016dcv}.  Only MACHOS which  are not part of the baryonic matter, such a PBH, could be a major DM component. It has been suggested~\cite{Bird:2016dcv} that the black holes recently detected by  the LIGO  collaboration~\cite{Abbott:2016blz} through gravitational waves emitted when two of them coalesced are DM PBH. However, because of their black hole nature,  many more limits apply to PBH than to generic MACHOs which exclude PBH from constituting the bulk of the DM for any mass, except maybe in a few windows at the boundaries between two different limits, if these limits are weakened (see e.g.~\cite{Carr:2016drx}). Two of these PBH mass windows are indicated in Fig.~\ref{Fig-0} with a dashed red line. For the 20 M$_\odot$ to 100 M$_\odot$ window to exist for PBH two independent bounds  should be somehow evaded or weakened~\cite{Carr:2016drx}
 (on the survival of a star cluster in the Eridanus II dwarf galaxy~\cite{Brandt:2016aco} and another, more model dependent,  on the effects  that X-rays emitted by gas accreted onto PBH could have on the CMB~\cite{Ricotti:2007au}).
  If LIGO has detected DM PBH,  then  the black hole mergers they observe should be spatially distributed  as the DM and not as the luminous matter. This will be tested as LIGO collects more merger events.
  
 The limits just presented and the lack of other candidates besides PBH and new elementary particles that can have the right relic abundance to be the DM,  constitute the only observational arguments we have in favor of elementary particles as DM  candidates.
 
The large scale structure of the Universe has shown that the bulk of the DM must be Cold or Warm (instead of Hot), i.e. either non-relativistic or becoming non-relativistic (instead of relativistic)  when galactic-size perturbations became encompassed by the growing horizon $\simeq ct$. This happened when the temperature of the Universe was $T \simeq$ keV. Particles that have a thermal spectrum have a characteristic energy of the order of the temperature $T$, thus are Cold DM if $m > $ keV, Warm DM if $m \simeq$ keV and Hot DM if $m <$ keV.  The only particles in the Standard Model of elementary particles which are neutral and without strong interactions as required for the DM,   are neutrinos and they are Hot DM because their masses are much smaller than a keV and they have a thermal spectrum in the early Universe (they must be in thermal equilibrium during BBN). 

There are no Cold or Warm DM candidates in the Standard Model. Thus particle DM requires physics beyond the Standard Model. A plethora of DM particle candidates have been proposed. The mass range of several of them are shown in Fig.~\ref{Fig-0}.   The lower limit on the DM particle mass is at least 10$^{-31}$ GeV since there is a concrete particle candidate proposed with this mass, the ``Fuzzy DM". This is a boson with a de Broglie wavelength of 1 kpc~\cite{fuzzy-DM}.

Models for beyond the Standard Model physics at the electroweak scale, such as supersymmetric models, usually predict  Weakly Interacting Massive Particles (WIMPs) as DM candidates, although these are by no means the only WIMP models. WIMPs are characterized by interacting with Standard Model particles with cross sections typical of weak interactions and have a mass in the 1 GeV-100 TeV range. Several independent DM searches have shown hints of a DM detection pointing to WIMPs in the lower portion of this mass range, 1 GeV  to 10's of GeV, which we call light WIMPs.

WIMPs became  preferred DM candidates early on, in the 1980's, because cross sections of weak order  expected in most beyond the Standard Model particle models  guarantee the right relic abundance for ``thermal" WIMPs. This is sometimes called the ``WIMP miracle". We usually characterize WIMP candidates according to how they are produced as ``thermal" or ``non-thermal" relics (see e.g.~\cite{Gelmini:2015zpa}).  Thermal relics are produced in the early Universe via interactions with the thermal bath, reach equilibrium with the bath and then ``decouple" or ``freeze-out" when their interactions cannot keep up with the expansion of the Universe (i.e. they become too rare to interact).
 After freeze-out  the number of WIMPs per comoving volume remains constant. 
  Non-thermal DM relics are all those not produced in this way (see e.g.~\cite{Baer:2014eja}). For example, they could be produced via the decay of other particles, which themselves may or may not have a thermal abundance.
 
 The relic density of thermal WIMPs, in the absence of a DM particle-antiparticle asymmetry, is inversely proportional to their annihilation cross section at freeze-out, actually to $\left<\sigma v\right>$, the average over the WIMP momentum distribution of the annihilation cross section times the relative WIMP speed $v$. If the Universe is radiation dominated at freeze-out and no significant change in entropy of matter and radiation occurs after,  one finds that  if $\left<\sigma v\right> = 3 \times10^{-26}$ cm$^3$/s for s-wave annihilation (for which $\left<\sigma v\right>$ is $v$ independent) or  $\left<\sigma v\right> = 6 \times10^{-26}$ cm$^3$/s for p-wave annihilation (for which $\left<\sigma v\right>
\sim v^2$)  thermal WIMPs freeze-out when $T\simeq m/20$ with the right DM density. These are cross sections of weak order for WIMPs in their characteristic mass range. However, notice that the thermal WIMPs freeze-out happens  at temperatures larger than 4 MeV for all WIMPs with mass $m > 80$ MeV. Since we do not know the history of the Universe before its temperature was 4 MeV cosmological assumptions  must be made to compute the relic abundance of these WIMPs, which may be wrong. Modifications of the standard cosmological assumptions (or a DM particle-antiparticle asymmetry) can result in vastly different predictions of the relation between WIMP relic density and cross section (see e.g.~\cite{Gelmini:2006pw})

The efforts to determine what the DM consists of are numerous.  They involve particle physics, cosmology and astrophysics. WIMPs are searched for in direct detection experiments, indirect detection experiments and  at colliders. 

Direct detection experiments try to detect the energy deposited in a detector by collisions of WIMPs in the dark halo on our galaxy. Direct detection is, therefore,  subject to uncertainties in the local dark halo characteristics.

Indirect detection looks for annihilation or decay products either of WIMPs in the dark halos of the Milky Way and other galaxies, in particular photons or anomalous cosmic rays, such as positrons and antiprotons, which do not come from astrophysical sources or of WIMPs  captured by and accumulated within the Sun or Earth from which only neutrinos can escape.  

A clear caveat to indirect detection is that the DM may not annihilate in present times, or decay. 
The DM does not annihilate at present if it consists only  of particles which are different from their antiparticles. This can happen with ``Asymmetric DM". We owe our very existence to a particle-antiparticle asymmetry in protons and neutrons versus their antiparticles. It insures that after all particle-antiparticle pairs annihilate in 
the early Universe,  the excess of particles survive. This explains why atomic matter only consists of matter and not antimatter at present. In Asymmetric DM models, considered also since the 1980's~\cite{Nussinov:1985xr},  a similar production mechanism is assumed for the DM (see e.g.~\cite{Petraki:2013wwa}).  

In most DM particle models the DM is stable because it is the lightest carrying a conserved quantum number, thus it cannot decay. For example, in supersymmetric models each Standard Model particle has a supersymmetric partner whose spin differs by $1/2$,  and there is a parity,  ``R-parity" under which all supersymmetric partners are odd and all the Standard Model particles are even. R-parity conservation guarantees that the lightest supersymmetric particle is stable (also that supersymmetric particles can only be created in pairs from Standard Model particles).

Indirect detection requires a good understanding of astrophysical backgrounds and the expected signal is  subject to uncertainties in the characteristics of the dark halos.
  
  At colliders, in particular the LHC (Large Hadron Collider),
  DM particles  produced in the collisions would escape from the detectors without leaving any trace. Because of energy and momentum conservation in each collision, the signal of DM particles is missing energy and momentum in the collision products.   An advantage of lepton colliders, e.g. electron-positron colliders, is that the total  missing energy and momentum can be measured, but there is no high energy collider of this type operating at the moment. In hadron colliders, such as the LHC in which protons collide with protons, the actual colliding particles are  the hadron constituents, or partons, the quarks and gluons. Each parton carries a fraction of the incoming hadron momentum. Because the initial momentum along the hadronic beam axis of the colliding partons is not known, the total amount of missing energy and momentum cannot be determined. However, the initial partonic momentum transverse to the beam axis is zero, so any net momentum of the collision products is an indication of missing transverse momentum. This is usually called Missing Transverse Energy (MET), although the name is correct only if the escaping particles are massless. MET is the signal of neutrinos produced in the collisions, whose mass is completely negligible with respect to their energies. Neutrinos were, in fact,  discovered using their signature of missing energy and momentum. In 1930 Wolfgang Pauli postulated their existence as a way of preserving the conservation of energy and momentum in beta decays.
    
 Most WIMPs produced at colliders are relativistic and a particle traveling close to the velocity of light escapes the detectors in about 100ns (30 meters).  Thus,  one should keep in mind that a  MET signal produced by a particle with lifetime $\simeq 100$ ns cannot be distinguished at colliders from one which lives longer than the lifetime of the Universe, $4 \times 10^{17}$ s, as required for DM particles.  Thus, even if a signal is found at the LHC for a potential DM particle, its presence  in the Universe constituting the DM at present will need to be confirmed by other means. 

Other caveats to the LHC searches are that DM particles with mass larger than a few TeV would be too heavy to be produced and  that these searches are relatively insensitive to DM interacting only with leptons.

\section{Light WIMPs in Direct Detection}  

The expected flux of  WIMPs passing through a detector on Earth is very large. Considering that the maximum local WIMP number density is $n=\rho_{\rm DM}/m$, where  $\rho_{DM} \simeq 0.3$  GeV/cm$^3$ is the local  DM density and $m$ is the WIMP mass,  and that the characteristic WIMP speed is $v\simeq 10^{-3}$ in the galaxy, the WIMP flux is  $n v \simeq 10^7$(GeV/$m$)/cm$^2$s (notice that  all along we use natural units in which c$=$1 and $\hbar=$1, thus e.g.  1s $\simeq 3 \times 10^{10}$cm). What makes the direct detection of WIMPs very difficult is not that their flux is small but that they interact with Standard Model particles very rarely and, when they do, deposit a small amount of energy, typically 1 to 10's of keV.  Present upper limits on the interaction rate of e.g.  WIMPs with a spin-independent interaction (Eqs.~\ref{dsigmadE} and  \ref{sigma-SI})  are at present  below 1 event/(100 kg year) for $m=$ 60 GeV, thus requiring ton scale detectors, and  only below 1 event/(kg day) for light WIMPs. This allows small detectors of kg size with low enough energy threshold to be relevant  for light WIMPs. 

The small rates and deposited energies force the experiments to have a extremely good control of backgrounds.  The detectors must be located underground, either under mountains or in mines, to decrease the background due to the interaction of cosmic rays in the atmosphere.

Most direct searches are non-directional, although some, still in the stage of development, attempt to measure the recoil direction besides the energy (see e.g.~\cite{Mayet:2016zxu}) and they are not relevant for our main topic.

\subsection{The expected rate in direct detection experiments} 

The recoil rate expected (in non-directional detectors) is given by the local WIMP flux, $n v$, 
times the number of target nuclei in the detector, times the scattering cross section, integrated over the local WIMP velocity distribution with respect to Earth. 
 The expected differential recoil rate, usually given in units of events/(kg of detector)/(keV of recoil energy)/day,  is
\begin{equation}
\frac{dR}{dE_R} =  \sum_T \frac{dR_{T}}{dE_R}=\sum_T \int_{v>v_{\rm min}}  \frac{C_T}{M_T}
\times \frac{d\sigma_T}{dE_R}  \times n  v f(\vec{v},t)~ d^3v.
\label{dRdE}
\end{equation}
Here $E_R$ is the recoil energy of the target nucleus, $M_T$ is its mass, $T$ denotes each target nuclide (elements and isotopes), $C_T$ is the mass fraction of nuclide $T$ in the detector, thus   $C_T/M_T$ is the number of targets $T$ per unit mass of detector, $d\sigma_T/dE_R$ is the differential WIMP-nucleus scattering cross section and $n= \rho/m$ is the local WIMP number density. The local WIMP density  $\rho$ coincides with the local DM density,  $\rho = \rho_{\rm DM}$, only if the WIMP in question constitutes all  of the DM. If instead  it constitutes only a fraction {\it r} $<1$ of the total, then $\rho =$ {\it r}$\rho_{DM}$. Besides, $\rho_{\rm DM}$ itself is not known beyond roughly a factor of 2 (see below). Thus  $\rho$ may actually be larger than the WIMP density assumed in a particular analysis by a factor {\it r} $>1$.  Since $\rho$ is a multiplicative factor in the rate of any direct detection experiment, its actual value does not affect the comparison of the results of direct detection experiments among themselves, but it affects the determination of the cross section from direct detection searches and the comparison of the results of direct searches with  indirect and collider searches, which have a different dependence on $\rho$.

The local DM density
$\rho_{\rm DM}$ and  the local WIMP velocity distribution with respect to Earth, $f(\vec{v},t)$, depend on the dark halo model adopted.  $f(\vec{v},t)$ depends on the time $t$ because of Earth's rotation around the Sun. The integral in velocity is over all WIMPs with speeds larger than  the minimum speed $v_{\rm min}$ a WIMP must have  to communicate a recoil energy $E_R$ to the target  nuclide $T$. 

The typical momentum transfer magnitude in a WIMP-nucleus collision is $q = |{\vec{q}}| \simeq \mu_T v \simeq$ O(MeV), where 
$\mu_T=m M_T/(m+M_T)$ is the WIMP-nucleus reduced mass. It is easy to check that $q \simeq$ MeV$\left({m}/{{\rm GeV}}\right)$ for $m\ll M_T$,  and $q \simeq A_T$ MeV for  $m\gg M_T$, where $A_T$ is the nuclear mass number of the target nuclide. The nuclear radius is $R_{\rm Nucleus} \simeq1.25~ {\rm fm} ~A_T^{1/3}$,  and $A_T^{1/3}$ is a number of about 3 to 5 for most nuclei.  Thus typically 
\begin{equation}
\label{coherence}
q < \frac{1}{R_{\rm Nucleus}}  \simeq {\rm MeV} \left({160}/{A_T^{1/3}}\right),
\label{q}
\end{equation}
what means that
WIMPs interact coherently with nuclei. If  $1/q \gg R_{\rm Nucleus}$ the nucleus interacts like a point-like particle, and for larger values of $q$ the loss of complete coherence is taken into account by a  nuclear form factor $F^2(E_R)$.

For contact interactions in the non-relativistic limit $v \to 0$ the differential cross section has the form 
\begin{equation}
\frac{d\sigma_T}{dE_R}=  \frac{\sigma_T(E_R)}{E_{max}} = \sigma_T(E_R)~\frac{M_T}{ 2\mu_T^2 v^2}.
\label{dsigmadE}
\end{equation}
for momentum transfer and velocity-independent interaction operators.  There are only two types of interactions of this kind: ``spin-independent" (SI) contact  interactions, due to a scalar or vector heavy mediator, and  ``spin-dependent" (SD) contact interactions, due to an axial vector heavy mediator. Heavy means here that the mediator mass is much larger than the momentum transfer in the scattering $q \simeq$ O(MeV). 

For SI interactions the DM particle couples to the nuclear density, thus
$\sigma^{SI}_T(E_R)$=$\sigma_{T0} F_T^2(E_R)$, where
\begin{equation}
\sigma_{T0} = \bigl[ Z_T + (A_T-Z_T) (f_n/f_p) \bigr]^2
(\mu_T^2/\mu_{\rm p}^2)\sigma_{\rm p},
\label{sigma-SI}
\end{equation}
and the nuclear form factor $F_T^2(E_R)$ is normalized to $F_T^2(0)$ = 1 (see e.g.~\cite{Jungman:1995df}). Here, $\mu_{\rm p}$ is the WIMP-proton reduced mass, $\sigma_{\rm p}$ is the WIMP-proton cross section, $f_n$ and $f_p$ are the WIMP couplings to neutrons and protons respectively, $Z_T$ is the number of protons and $(A_T-Z_T)$ is the number neutrons in the target $T$. If the WIMP couples equally to protons and neutrons, $f_n/f_ p=$ 1 (isospin-conserving coupling), the SI cross section is proportional to the square of the mass number, $\sigma_{T0} =A_T^2 (\mu_T^2/\mu_{\rm p}^2) \sigma_{\rm p}$,  but there is no reason for isospin to be conserved in the WIMP couplings.  Some values of the $f_n/f_ p$ ratio make the WIMP coupling to particular nuclei very small. Because heavier nuclei are more neutron rich than lighter nuclei, the ratio  $(f_n/f_p) = -Z_T/ (A_T -Z_T)$ cancels the WIMP coupling with the particular nuclide with atomic and mass numbers $Z_T$ and $A_T$ (and also changes the couplings to all other elements).  Notice that no choice of $f_n/f_ p$ can make  the coupling  with an  element zero because there is a natural isotopic composition, with nuclides of the same $Z_T$ and 
different $A_T$.  The so called  ``Isospin-Violating DM"~\cite{IsospinViolating} with  $f_n/f_p = -0.7$ minimizes the coupling with xenon (thus weakening the limits of XENON10, XENON100 and LUX, some of the strongest  at present).   Instead, $f_n/f_p = -0.8$ reduces maximally the coupling to germanium~\cite{Gelmini:2014psa} (thus weakening preferentially the limits of CDMS and SuperCDMS, also among the strongest). WIMP particle models in which $f_n/f_p$ can have these negative values have been proposed (see e.g.~\cite{DelNobile:2011je}).

For SD interactions the DM particle couples to the nuclear spin density, leading to 
\begin{equation}
\sigma^{SD}_T(E_R)= 32\mu_T^2G_F^2 \left[(J_T+1)/J_T\right] \bigl[ \langle S_{\rm p}\rangle 
a_{\rm p} + \langle S_{\rm n}\rangle 
a_{\rm n} \bigr]^2 F^2_{SD}(E_R),
\label{sigma-SD}
\end{equation}
(see e.g.~\cite{Jungman:1995df}). Here, $J_T$ is the nuclear spin,  $a_{\rm p,n}$ are the WIMP couplings to p and n and $F^2_{SD}(E_R)$ is the  nuclear form factor, with
$F^2_{SD}(0)=$ 1. $\langle S_{\rm p,n}\rangle$, the expectation values of the proton and neutron spin content in the target nucleus, are numbers $\lsim$ O(1) that can differ easily by factors of 2 or more in different nuclear models.

There are many other types of possible WIMP-nucleus interactions besides the two mentioned, and many of them have been considered in recent years to try to accommodate different DM hints  in direct or indirect searches. All other interaction operators contain extra powers of the momentum transfer $q$ or the WIMP velocity (\cite{Fitzpatrick:2012ix,  Barello:2014uda} list all possible operators in the non-relativistic limit up to  $O(q^2)$). For example,  a pseudo-scalar mediator produces an interaction dependent on the component of the nuclear and WIMP spins in the direction of  $\vec{q}$, with two extra powers of $q=|\vec{q}|$. Moreover, the mediators can be either heavy or light with respect to $q$.

An interesting possibility explored recently  is that of neutral DM particle candidates which interact with photons through higher electromagnetic moments. 
For fermionic DM, the most studied candidates with electromagnetic moments are WIMPs with magnetic or electric dipole moments, e.g. in~\cite{Pospelov:2000bq, Sigurdson:2004zp, Chang:2010en, DelNobile:2013cva, Barello:2014uda}, which have the lowest order electromagnetic interactions possible, given by dimension five effective operators suppressed  by one power of a new physics mass scale $\Lambda$ in the denominator. Magnetic and electric dipole moments vanish for Majorana fermions (although nonzero transition moments are possible), and the only possible electromagnetic moment is an anapole moment. The interaction in this case is described by an effective dimension six operator weighted down by two powers of a new physics mass scale $\Lambda$ in the denominator. Anapole moment DM (Anapole DM) has been studied in the context of direct detection e.g. in~\cite{Pospelov:2000bq, Ho:2012bg, DelNobile:2013cva}. In this case the scattering cross section is~\cite{DelNobile:2013cva}
\begin{equation}
\label{ADM}
\hspace{-1.5 cm} \frac{d\sigma_T}{d E_R} = \sigma^A_{\rm ref} \frac{M_T}{\mu_N^2} \frac{v_{\rm min}^2}{v^2} \left[ Z_T^2 \left( \frac{v^2}{v_{\rm min}^2} - 1 \right) F_{{\rm E}, T}^2(q^2) + 2 \frac{\lambda_T^2}{\lambda_N^2} \frac{\mu_T^2}{m_N^2} \left( \frac{J_T + 1}{3 J_T} \right) F_{{\rm M}, T}^2(q^2) \right],
\end{equation}
where $q^2$ is the square of the momentum transfer, $M_T$ is the nuclear mass, $m_N$ is the nucleon mass, $\mu_N$ and $\mu_T$ are the WIMP-nucleon and WIMP-nucleus reduced masses, respectively, $\lambda_N = e / 2 m_N$ is the nuclear magneton, $\lambda_T$ is the nuclear magnetic moment, $J_T$ is the nuclear spin, and the reference cross section is defined as
$\sigma^A_{\rm ref} \equiv 2 \mu_N^2 \, \alpha g^2 / \Lambda^4$
($\alpha = e^2 / 4 \pi \simeq 1/137$ is the fine structure constant and  $g$ is a dimensionless coupling constant).  Notice that the Anapole DM cross section is very different from the usual SI or SD cross sections, Eqs.~(\ref{dsigmadE}), (\ref{sigma-SI}) and (\ref{sigma-SD}).  In particular, the $v$ dependence of the cross section does not factorize as $1/v^2$. The first term in Eq.~(\ref{ADM}) corresponds to the interaction with the nuclear charge $Z_T$ and the second with the nuclear magnetic field. Each has its own nuclear form factor, $F_{{\rm E}, T}^2(q^2)= (4\pi / Z^2) F_{\rm L}^2(q^2)$ and $F_{{\rm M}, T}^2(q^2) = (3J_T / J_T+1) (8\pi m_N^2 / q^2) (\lambda_N^2 /\lambda_T^2) F_{\rm T}^2(q^2)$, respectively. $F_{\rm L}^2$ and $F_{\rm T}^2$ are the standard longitudinal and transverse nuclear form factors but normalized to 1 at $q^2=0$ (see~\cite{DelNobile:2013cva} for details).

Even when the DM elementary particle model is stablished there are uncertainties in the scattering cross section. Starting from the fundamental interactions with which a particular WIMP candidate couples to quarks, there are uncertainties on how to pass from quarks to protons and neutrons, and then to nuclei. Moreover, each type of interaction requires its own nuclear form factor, most of which are poorly known. 

With the contact differential cross-section of ~Eq.~(\ref{dsigmadE}),  Eq.~(\ref{dRdE}) becomes
\begin{equation}
\frac{dR}{dE_R} =
\sum_T \frac{\sigma_T(E_R)}{2 m \mu_T^2}  \rho \eta(v_{\rm min}, t),
\label{dRdE-2}
\end{equation}
where the halo function $\eta(v_{\rm min}, t)$ is defined as
\begin{equation}
\eta(v_{\rm min}, t) = \int_{v>v_{\rm min}}
\frac{f(\vec{v},t)}{v} d^3v.
\label{eta}
\end{equation}
Notice that the factor $\rho \eta(v_{\rm min}, t)$ includes all the dependence of the rate, Eq.~\ref{dRdE-2},
on the dark halo model for any detector. The minimum speed $v_{\rm min}$  depends on the collision being elastic or inelastic. In some particle models a DM particle of mass $m$ may collide inelastically, producing a different state with mass $m' = m + \delta$~\cite{TuckerSmith:2001hy}, while the elastic scattering is either forbidden or suppressed. The mass difference $\delta$ can either be positive (for ``inelastic DM", iDM, with endothermic scattering)~\cite{TuckerSmith:2001hy} or negative (for ``exothermic DM", exoDM, with exothermic scattering)~\cite{Graham:2010ca}.
For elastic collisions $\delta=0$.   For $\left| \delta \right| \ll m$ (actually $\mu_T|\delta|/m^2 \ll 1$)
\begin{equation}
v_{\rm min}=\ \left|  \sqrt{\frac{M_T E_R}{2\mu_T^2}}  + (\frac{\delta}{\sqrt{2M_T E_R}}) \right|.
\label{vmin}
\end{equation}

The differential recoil rate $dR_{T}/dE_R$  is not directly experimentally accessible because of energy-dependent experimental efficiencies and  energy resolution functions. The recoil energy $E_R$ is not directly measured. What is measured is a proxy $E'$ for $E_R$ (many times $E'$ is only a fraction of the recoil energy, the fraction going to scintillation or ionization). The observable differential rate is
\begin{equation}
\label{Obs-rate}
\frac{dR}{dE'} = \epsilon(E') \, \int_0^\infty dE_R \, \sum_{T}  G_{T}(E_R,E') \, \frac{dR_{T}}{dE_R}.
\end{equation}
Here $E'$ is the detected energy, often quoted in keVee (keV electron-equivalent) or in photoelectrons, $\epsilon(E')$ is a counting efficiency or cut acceptance, and $G_T(E_R, E')$  is a (target nuclide and detector dependent) effective energy resolution function that gives the probability that a recoil energy $E_R$ is measured as $E'$.  $G_T$ incorporates the mean value of the observed energy, e.g. $\langle E' \rangle = Q_T E_R$ where  $Q_T(E_R)$ is an energy dependent ``quenching factor",  and the energy resolution $\sigma_{E_R}(E')$. These functions must be measured (sometimes $\sigma_{E_R}(E')$ is computed).

WIMP interactions in crystals produce mostly phonons. Only a fraction $Q_T$ of the recoil energy goes on average into ionization or scintillation.  For example, $Q_{\rm Ge} \simeq$ 0.3, $Q_{\rm Si} \simeq$ 0.25, $Q_{\rm Na} \simeq$ 0.3, and $Q_{\rm I} \simeq$ 0.09. In noble gases such as Xe,  a similar factor $L_{eff}$ measures the scintillation efficiency of a WIMP relative to a photon. There are large experimental uncertainties in the determination of these parameters at low energies.

 Eqs.~(\ref{dRdE}) and (\ref{Obs-rate}) show that three elements enter into the observable rate in  direct detection experiments: 1) the detector response, 2) the particle physics input, given by the cross section and mass of the DM candidate, and 3) the local dark halo model. 
%
\begin{figure}[h]
\centerline
{\includegraphics[width=0.70\textwidth]{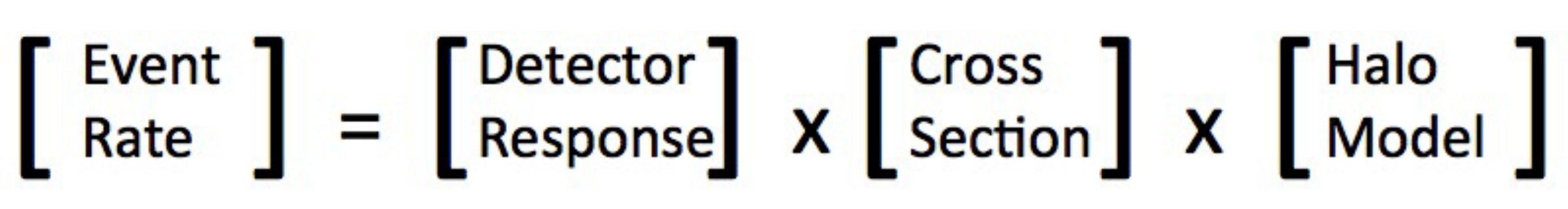}}
\end{figure}

\noindent There are uncertainties in all of these elements, which must be taken into account when comparing the results of different direct DM detection experiments.

\subsection{Halo-Dependent and Halo-Independent direct detection data comparison methods}

In direct searches without directionality the unmistakable signature of DM is an annual modulation of the rate, due to the variation in the velocity of the WIMPs with respect to Earth as Earth rotates around the  Sun~\cite{Drukier:1986tm}. 

A simple model which captures the bulk of the known characteristics of the dark halo is the Standard Halo Model (SHM), in which $\rho_{\rm DM} = $ 0.3 GeV/cm$^3$  and the local DM velocity distribution with respect to the galaxy is a Maxwell-Boltzmann with zero average  and dispersion $ v_0$, truncated at the local escape speed $v_{\rm esc}$  (see e.g.~\cite{Savage:2008er}). The  annual average velocity of WIMPs with respect to Earth is thus,  $-\vec{v}_\odot$, where  $\vec{v}_\odot$ is the velocity of the Sun with respect to the galaxy. Some usual values  for these parameters are $v_\odot=232$ km/s, ${v}_0=220$ km/s, and ${v}_{\rm esc}=533$ km/s, but there are large uncertainties (e.g. values  of  $v_{\rm esc}$ between 500  and 650 km/s can be found in the literature). 

In the SHM, disregarding gravitational focussing effects due to the Sun, the maximum average velocity of WIMPs with respect to Earth happens at the end of May-beginning of June (see e.g.~\cite{Bozorgnia:2012eg}) and the minimum six months later (not exactly, because of the small ellipticity of Earth's orbit). However, for WIMPs with low speeds, below 200 km/s, gravitational focussing by the Sun is important  and changes these dates~\cite{Lee:2013wza}.

  We expect the actual halo to deviate  from the simplistic  SHM model. There are large uncertainties in the measured value of the local dark halo density. A comprehensive compilation of measurements up to 2014~\cite{Read:2014qva} shows that the 1$\sigma$ range of most determinations since 2010 are included in the 0.2 to 0.6 GeV/cm$^3$ range.  A more recent measurement using the rotation curve of the galaxy and assuming spherical symmetry  of the dark halo~\cite{Pato:2015dua}  finds $\rho_{\rm DM}= 0.420^{+ 0.019}_{-0.021} (2\sigma) \pm 0.026$ GeV/cm$^3$ where the second error is the standard deviation of the best-fit values over all visible matter models used in the paper (and there is a small variation of this range depending on the dark halo profile assumed).
 The local DM density and velocity distribution could also be affected by local substructure in the dark halo, e.g. if Earth is within a DM clump, which is unlikely~\cite{Vogelsberger}, or in a DM stream~\cite{Purcell:2012sh}, or if there is a dark disk in our galaxy~\cite{Read:2009iv}.  The DM of the leading arm of the Sagittarius Stream, tidally stripped from the Sagittarius Dwarf Galaxy, could be passing through the Solar system, perpendicularly to the galactic disk~\cite{Purcell:2012sh} and tidal disruption of earlier substructure have created ``debris flows", which are spatially homogeneous structures in velocity~\cite{Lisanti:2011as}.
 
To avoid astrophysical uncertainties in the modeling of the local dark halo, attempts have been made to compare sets of direct detection data without assuming any model for the dark halo, in a ``Halo-Independent" manner, as opposed to the usual ``Halo-Dependent" manner.

The   Halo-Dependent data comparison  method, used since the inception of direct 
detection~\cite{Ahlen:1987mn}, fixes the three aforementioned  elements of the observable rate, usually assuming the SHM for the galactic halo, except for the WIMP mass $m$ and a reference  cross section parameter $\sigma_{\rm ref}$ extracted from the cross section (e.g. $\sigma_{\rm ref}=\sigma_p$ for SI interactions) which parameterizes the magnitude of the cross section. Data are plotted in the $m, \sigma_{\rm ref}$ plane (as in Figs. 1.a, 2.a and 3.a)
assuming  a particular value, usually 0.3 GeV/cm$^3$, for the local WIMP density $\rho$. If  instead $\rho$ differs from this assumed value by a factor {\it r}, then  $\sigma_{\rm ref}$  should be replaced by {\it r}$\sigma_{\rm ref}$. 

In the  ``Halo-Independent" data comparison method~\cite{Fox:2010bz} elements 1) and 2) of the observable rate (the detector response and the particle physics input) are fixed, except again for a  parameter $\sigma_{\rm ref}$ extracted from the cross section, but no assumption is made about the element 3), the dark halo model. This method was initially developed for  SI  WIMP-nucleus interactions~\cite{Fox:2010bz, Frandsen:2011gi, Gondolo:2012rs} and only in~\cite{DelNobile:2013cva} extended to any other type of interaction. In the formulation of~\cite{Gondolo:2012rs} and~\cite{DelNobile:2013cva}, the rate in an observed energy interval $[E'_1,E'_2]$ for any type of WIMP-nucleus interaction is written as 
\begin{equation}
\label{HaloIndep-rate} 
  R_{[E'_1,E'_2]}= \int_{E'_1}^{E'_2} dE'~ {dR}/{dE'} =\int_0^\infty dv_{\rm min}  \mathcal{R}_{[E'_1,E'_2]}(v_{\rm min})\tilde\eta(v_{\rm min}, t),
  \label{HI-rate}
\end{equation}
  with a DM candidate and detector dependent response function $\mathcal{R}_{[E'_1,E'_2]}(v_{\rm min})$ and a function  $\tilde\eta(v_{\rm min},t)  \equiv (\sigma_{\rm ref}\rho/m) \eta(v_{\rm min},t)$, common to all experiments, that contains all the dependence  of the rate on the halo model~\cite{Gondolo:2012rs, DelNobile:2013cva} (the function $\eta(v_{\rm min},t)$ was defined in Eq.~(\ref{eta})). For each particular energy interval $[E'_1,E'_2]$ the response function $\mathcal{R}_{[E'_1,E'_2]}(v_{\rm min})$ is significantly non-zero only in a particular 
  $v_{\rm min}$ interval. 
  
  For a fixed WIMP mass $m$,  all rate measurements  and bounds can be mapped onto  the ${\rm v}_{min}, \tilde{\eta}$ plane.  To be compatible, experiments must measure the same  $\tilde{\eta}$ function. 
Due to the revolution of the Earth around the Sun,  $\tilde{\eta}({\rm v}_{min},t)$ has an annual modulation generally well approximated by the first two terms of a harmonic series,
\begin{equation}
\tilde\eta(v_{\rm min}, t)= \tilde\eta^{0}(v_{\rm min}) + \tilde\eta^{1}(v_{\rm min}) \cos(\omega (t - t_0)),
\label{eta-time} 
\end{equation}
where $t_0$ is the time of the maximum of the signal and $\omega = 2 \pi/$yr. The unmodulated and modulated components $\tilde{\eta}^0(v_{\rm min})$ and $\tilde{\eta}^1(v_{\rm min})$ are detector-independent quantities that must be common to all direct DM detection experiments and enter respectively in the definition of the unmodulated and modulated parts of the  rate of each experiment,
\begin{equation}
R_{[E'_1,E'_2]}(t) = R^0_{[E'_1,E'_2]} + R^1_{[E'_1,E'_2]} \cos(\omega (t - t_0)).
\label{Rt}
\end{equation}

  Notice that in Eq.~(\ref{HI-rate}) the independent variables are $v_{\rm min}$ and the observed energy $E'$. Taking $E_R$ as independent variable, as  done e.g. in~\cite{Fox:2010bz}, $v_{\rm min}$ is different for each target nuclide.  Notice that $E_R$ and $v_{\rm min}$ are exchangeable variables only for a single nuclide. In this case, the speed $v_{\rm min}$ is the minimum speed necessary for the incoming interacting DM particle to impart a recoil energy $E_R$ to a nucleus and, conversely, given an incoming WIMP speed $v=v_{\rm min}$, $E_R$ is the extremum recoil energy (maximum energy for elastic collisions, or either maximum or minimum for inelastic collisions) that the DM particle can impart to a nucleus.  When a target consists of multiple nuclides, a choice must be made between the two, $E_R$ and $v_{\rm min}$, as independent variable. Choosing $v_{\rm min}$ allows to incorporate any isotopic composition of the target by summing over target nuclide dependent $E_R(v_{\rm min})$ for fixed observed $E'$.

\subsection{Hints of Light WIMPs in direct detection experiments?}

There are many direct DM detection experiments that are either running, in construction or at the stage of research and development (see e.g.~\cite{Cushman:2013zza} and references therein).  They use different target materials (indicated here in parenthesis) and  detection strategies.  Three direct detection experiments, DAMA/LIBRA (NaI)~\cite{Bernabei:2013xsa}, 
CoGeNT (Ge)~\cite{Aalseth:2010vx, Aalseth:2011wp, Aalseth:2014eft} and CDMS-II-Si (Si)~\cite{Agnese:2013rvf},  have observed potential signals of DM. CRESST-II (CaWO$_4$)  with an upgraded detector no longer found in 2014~\cite{Angloher:2014myn} the unexplained  rate excess it had found in 2010~\cite{Angloher:2011uu}. All other direct detection searches, including LUX (Xe), XENON100 (Xe), XENON10 (Xe), CDMS-II-Ge (Ge), CDMSlite (Ge) and SuperCDMS (Ge), have produced only upper bounds on the interaction rate and on the annual modulation amplitude of a potential  WIMP signal (see e.g~\cite{DelNobile:2014sja} and references therein). 

 The DAMA/LIBRA  and earlier DAMA/NaI experiments of the same collaboration  (collectively referred to as DAMA  here), located at the Gran Sasso Laboratory, Italy,  have so far found in 19 years  of cumulative data, with an impressive exposure of 1.33 ton  year, an annual modulation in their 2 to 6 keVee bin (their threshold was 2 keVee) at the 9.3 $\sigma$ C.L.~\cite{Bernabei:2013xsa} with a period of 1 year and phase compatible with that expected from DM in the SHM.  
 Until 2003, due to theoretical prejudices, DAMA/NaI had cut the region of compatibility in its fits  to WIMP  masses $m >$ 30 GeV and by 2002 this region was excluded by EDELWEISS (Ge) and CDMS (Ge) data. Although the first proposed WIMP candidate was a light WIMP (a heavy neutrino of a few GeV studied by Lee and Weinberg in 1977~\cite{Lee:1977ua}, rejected by the first direct detection limits in 1987~\cite{Ahlen:1987mn}), the popularity of supersymmetric models in the 1990's led to a preference for heavier WIMP candidates. The lower limit of about 30 GeV on the mass of the lightest neutralino, the usual supersymmetric WIMP candidate, was used by DAMA/NaI when fitting their data until 2003 (when a model for a 6 GeV neutralino was proposed~\cite{Bottino:2003cz}).
 
   In 2004-2005 P. Gondolo and I~\cite{Gelmini:2004gm} proved that the DAMA/NaI modulation, interpreted as a signal of WIMPs in the SHM was still compatible with all the upper limits at the time for light WIMPs with SI interactions (and $f_n/f_p=1$), with $m =$ 5 to 9~GeV and WIMP-proton cross section $\sigma_p \simeq 10^{-40}$ cm$^2$. 
  However, the interest in light WIMPs did not start in earnest until 2008, when the DAMA/NaI annual modulation was confirmed by DAMA/LIBRA. Many papers reanalyzed the issue of compatibility of the DAMA data with all other negative searches at the time (see e.g.~\cite{Savage:2008er}) and found that it depended strongly on the possibility of ``channeling" of the recoiling target ion as estimated in~\cite{DAMA-chan}.   
  
   ``Channeling" and ``blocking" in crystals refer to the orientation dependence of ion penetration in crystals. 
  Channeling happens when an ion propagates inside a crystal along a symmetry axis or plane, so it gives all its energy to electrons instead of only a fraction.   In 2010 my collaborators and I~\cite{Bozorgnia:2010xy} recomputed
  an upper limit to the fraction of channelled recoiling ions, which originate from  lattice sites of the crystal, 
  and found that channeling was negligible (later confirmed experimentally~\cite{Collar:2013gu}). For ions ejected from lattice sites the  ``blocking" effect, namely the reduction  along symmetry axes and planes of the flux of ions due to the shadowing effect of the lattice atoms directly in front of the emitting lattice site, is very important~\cite{Bozorgnia:2010xy} and had not been taken into account in previous calculations.
  
  The interest in light WIMPs was reinforced in 2010 when CRESST-II~\cite{Angloher:2011uu} and CoGeNT found an unexplained  rate excess and later, in 2011, CoGeNT~\cite{Aalseth:2010vx} announced a hint of annual modulation (at  the $2\sigma$ and later smaller C.L.), all attributable to light WIMPs.
  
   CRESST-II (located also at the Gran Sasso Laboratory) with 730 kg day of exposure reported in 2011 to have found 670  events potentially  due to DM and, after considering all possible backgrounds plus a WIMP signal to explain those events (assuming SI interactions, $f_n/f_p=1$ and the SHM), found two regions in the $m, \sigma_p$ plane for which the background only hypothesis was rejected at more than 4$\sigma$, with best-fit WIMP mass of  25.3 GeV and 11.6 GeV respectively~\cite{Angloher:2011uu}.  However, subsequently, in 2014 and 2015~\cite{Angloher:2014myn}, the same collaboration with an upgraded detector did not confirm the existence of an excess of events over background  attributable to DM.
  
  The CoGeNT collaboration operating a single 440 g Ge detector at the Soudan Underground Laboratory, Minnesotta,  USA, with a low threshold, 0.4 keVee, found in 2010~\cite{Aalseth:2010vx} in 56 days of data an ``irreducible excess" of bulk-like events (backgrounds are mostly  expected at the surface) below 3 keVee, which could be fitted by taking into account backgrounds and a potential WIMP signal (again assuming SI interactions, $f_n/f_p=1$ and the SHM) in the mass range 7-11 GeV (with a $\chi^2$ per degree of freedom of 20.1/18). However, a fit with background only, and no WIMP signal, could be done with similar significance~\cite{Aalseth:2010vx}. In 2011, CoGeNT  announced with 15 months  (442 days) of data that the irreducible excess of events had a 17\% annual  modulation at the 2.8$\sigma$ C.L. compatible with that observed by DAMA~\cite{Aalseth:2011wp}. In subsequent years, 2013 and 2014, the original excess was reduced by CoGeNT's better understanding  of its backgrounds. After 3.4 y (at the end of the experiment) the CoGeNT data showed a preference for light WIMP recoils of very low significance, below the 2$\sigma$ C.L.~\cite{Aalseth:2014eft}.
  
  The CDMS-II collaboration operated from 2003 to 2008 an array of mostly Ge and  some Si detectors at the Soudan Underground Laboratory. The Ge CDMS-II  data  provided some of the most constraining upper limits, but in 2013 the collaboration published the analysis of  140.2 kg-days of data acquired from July 2007 to Sept. 2008 with their Si detectors. They reported having found 
  three WIMP-candidate events with recoil energies of 8.2, 9.5, and 12.3 keV, with  a 5.4\% probability of being produced by their known backgrounds. The highest likelihood for a WIMP with Si interactions and $f_n/f_p=1$, and assuming the SHM, was for a mass of 8.6 GeV. 
  
All other direct detection experiments besides DAMA, CoGeNT and CDMS-II-Si had only negative results.
Are (some of) these positive and negative results compatible with each other for some type of WIMP?

\begin{figure}[t]
\includegraphics[width=0.48\textwidth]{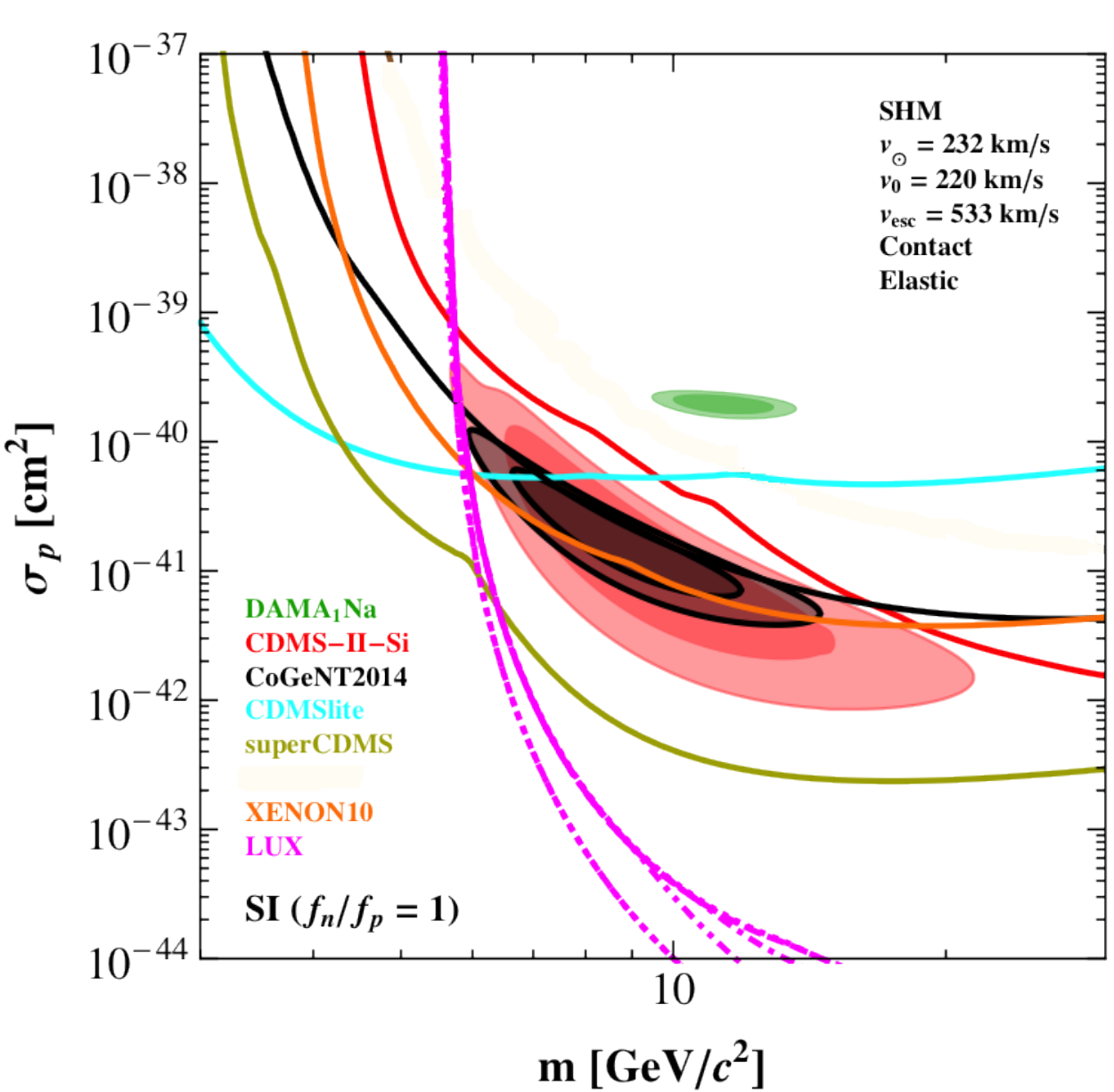}
~~\includegraphics[width=0.49\textwidth]{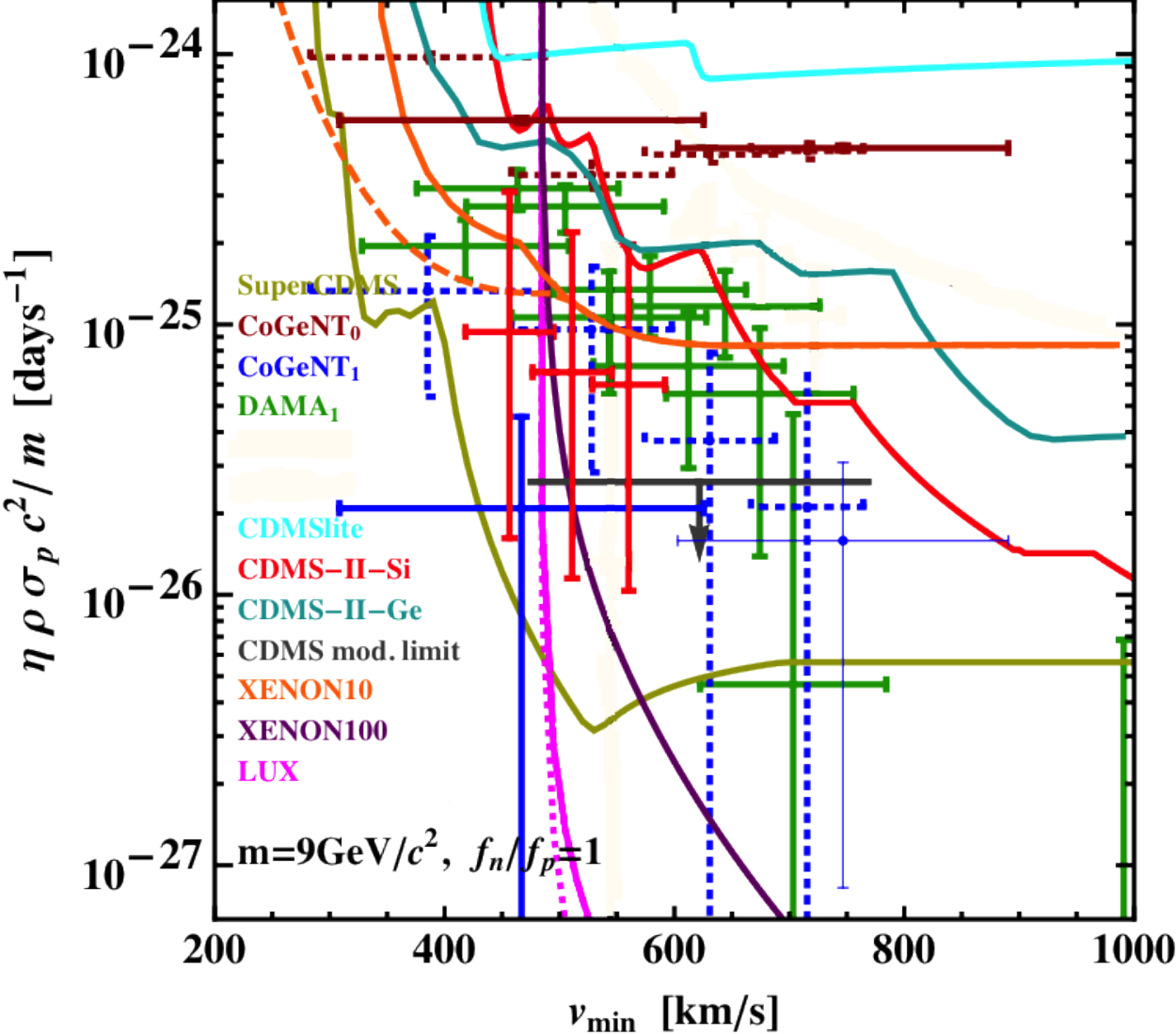}
\caption{\ref{Fig-1}.a (left) 90\%C.L. limits and 68\% and 90\%C.L. allowed regions assuming the SHM for SI $f_n/f_p$$=$1 interactions. \ref{Fig-1}.b (right) $1\sigma$ measurements of and 90\%C.L.  upper bounds on $\tilde\eta^0$ ($\tilde\eta$ average) for CDMS-II-Si (red crosses) and CoGeNT (brown crosses), and $\tilde\eta^1$ ($\tilde\eta$ annual modulation amplitude) for DAMA (green crosses) and CoGeNT (blue crosses) as function of $v_{min}$ for $m$= 9 GeV. See~\cite{Gelmini:2014psa} for details.}
\label{Fig-1}
\end{figure}

\begin{figure}[t]
{\includegraphics[width=0.48\textwidth]{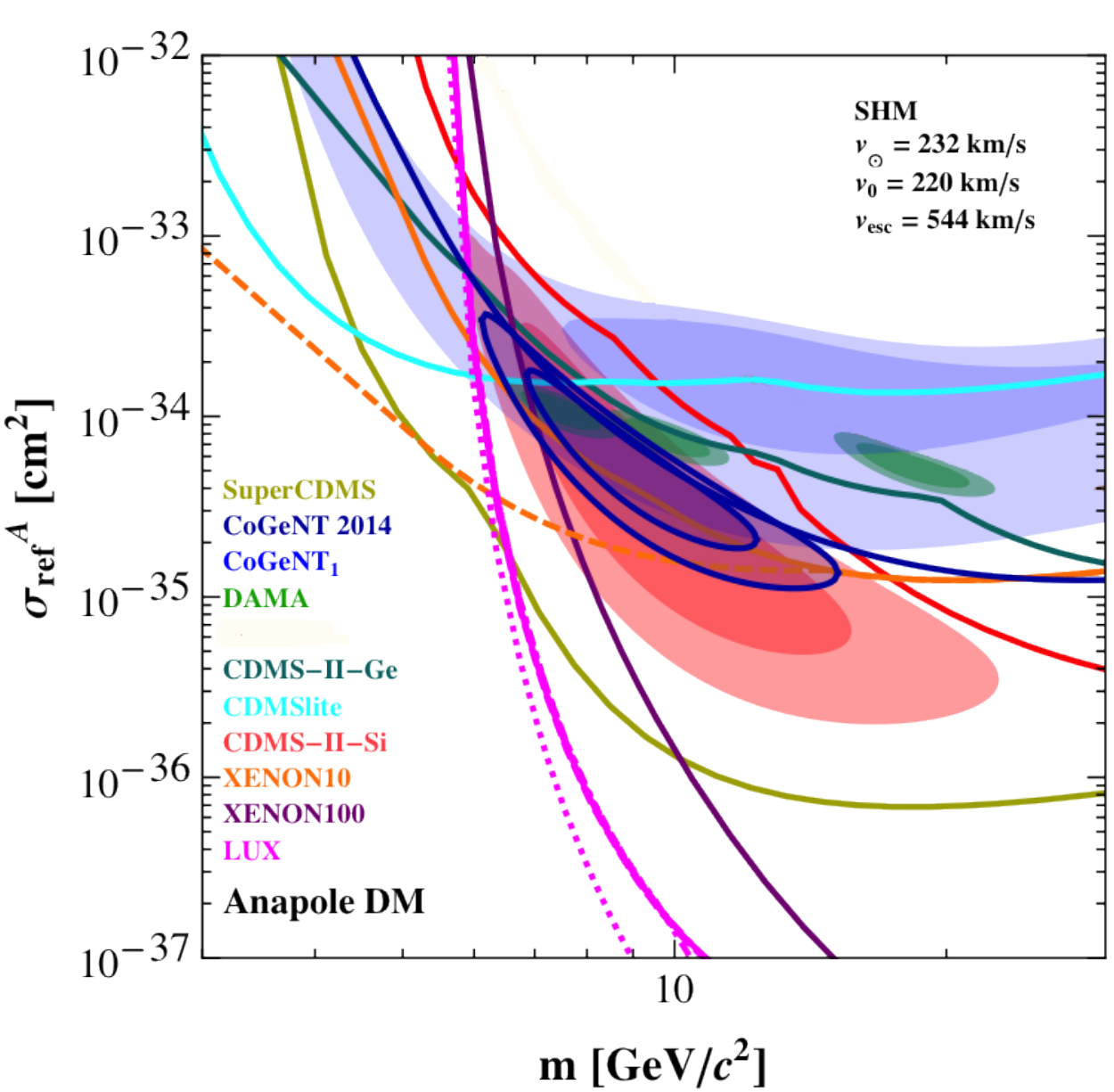}}
~~{\includegraphics[width=0.49\textwidth]{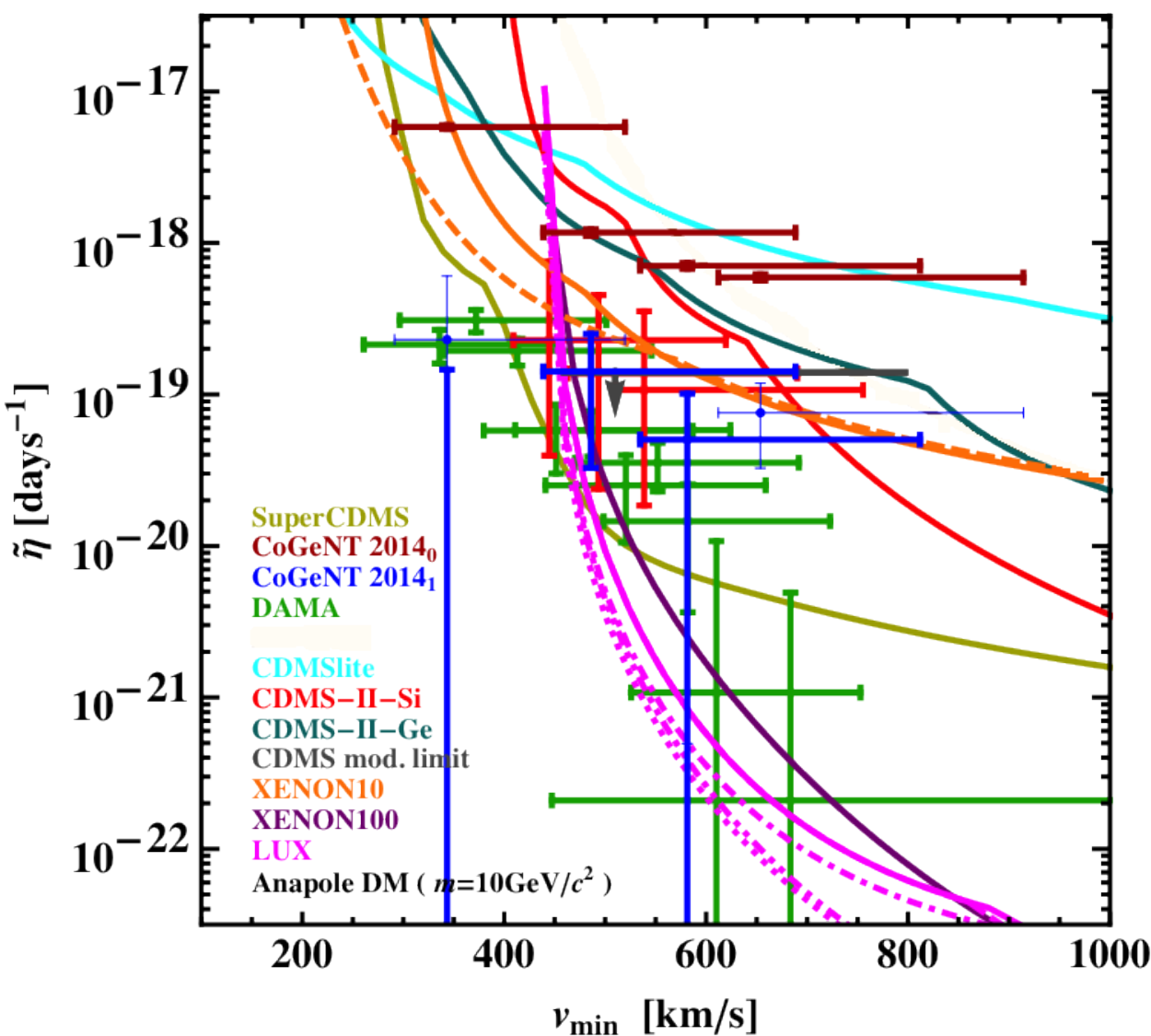}} 
\caption{\ref{Fig-2}.a (left) 90\%C.L. limits and $68\%$ and 90\%C.L. allowed regions for the DAMA (green) and CoGeNT 2011-2012 (light blue) modulation data (indicated by a subscript 1) and the CoGeNT 2014 (blue) and CDMS-II-Si (pink) unmodulated data, for Anapole DM,  assuming the SHM. The three (green) DAMA regions correspond to $Q_{\rm Na}$ equal to $0.45$ (left), $0.30$ (middle) and the energy dependent value from~\cite{Collar:2013gu} (right). \ref{Fig-2}.b (right) $1\sigma$ measurements of and 90\%C.L.  upper bounds on $\tilde\eta^0$ ($\tilde\eta$ average) for CDMS-II-Si (red crosses) and CoGeNT (brown crosses), and $\tilde\eta^1$ ($\tilde\eta$ annual modulation amplitude) for DAMA with $Q_{\rm Na} = 0.30$ (green crosses) and CoGeNT (blue crosses) as function of $v_{min}$ for Anapole DM with $m$= 10  GeV. See~\cite{DelNobile:2014eta} for details.}
\label{Fig-2}
\end{figure}

 The comparisons made by the experimental groups near always assume the SHM and either a SI WIMP-nucleus contact interaction with equal couplings to neutrons and protons ($f_n/f_p=1$) or  a SD WIMP-nucleus contact interaction. As shown in 
  Fig.~\ref{Fig-1}.a, for light WIMPs  with SI interactions with $f_n/f_p=1$ and the SHM the signal regions in the $m, \sigma_{\rm ref}$ plane are tantalizingly close, but they are excluded by several upper limits.
 The allowed regions and upper limits change with the type of WIMP-nucleus interactions. For example, for light WIMPs with anapole or magnetic dipole interactions,  the DAMA, CoGeNT and CDMS-II-Si  $m, \sigma_{\rm ref}$ regions overlap  (assuming the SHM and elastic interactions), thus all three  signals  could arise from the same DM candidate, if it were not that the regions are rejected by several upper limits, mostly  those of SuperCDMS and LUX~\cite{DelNobile:2014eta}. This is shown in Fig.~\ref{Fig-2}.a for Anapole DM.

\begin{figure}[t]
{\includegraphics[width=0.48\textwidth]{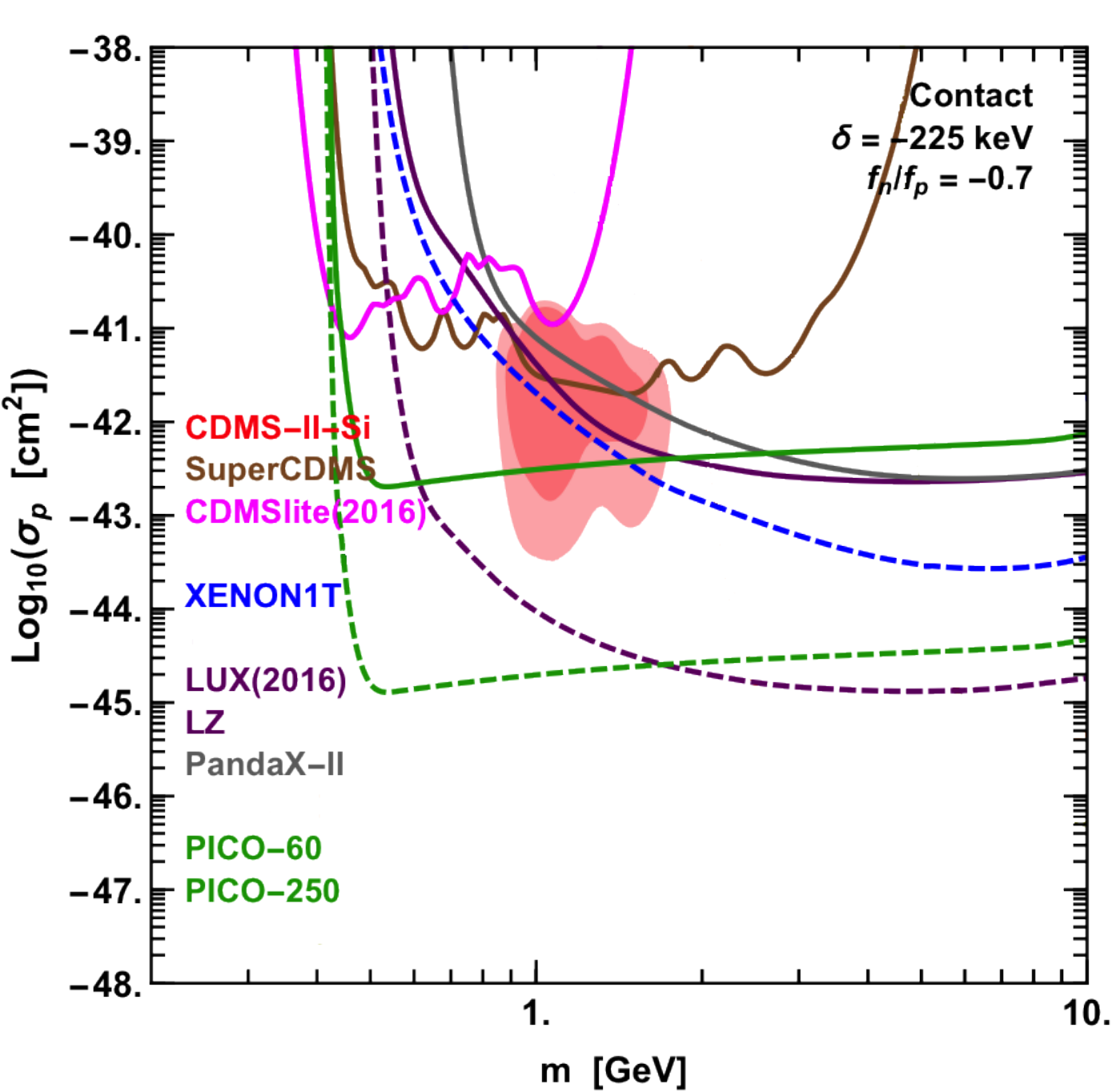}}
~~{\includegraphics[width=0.46\textwidth]{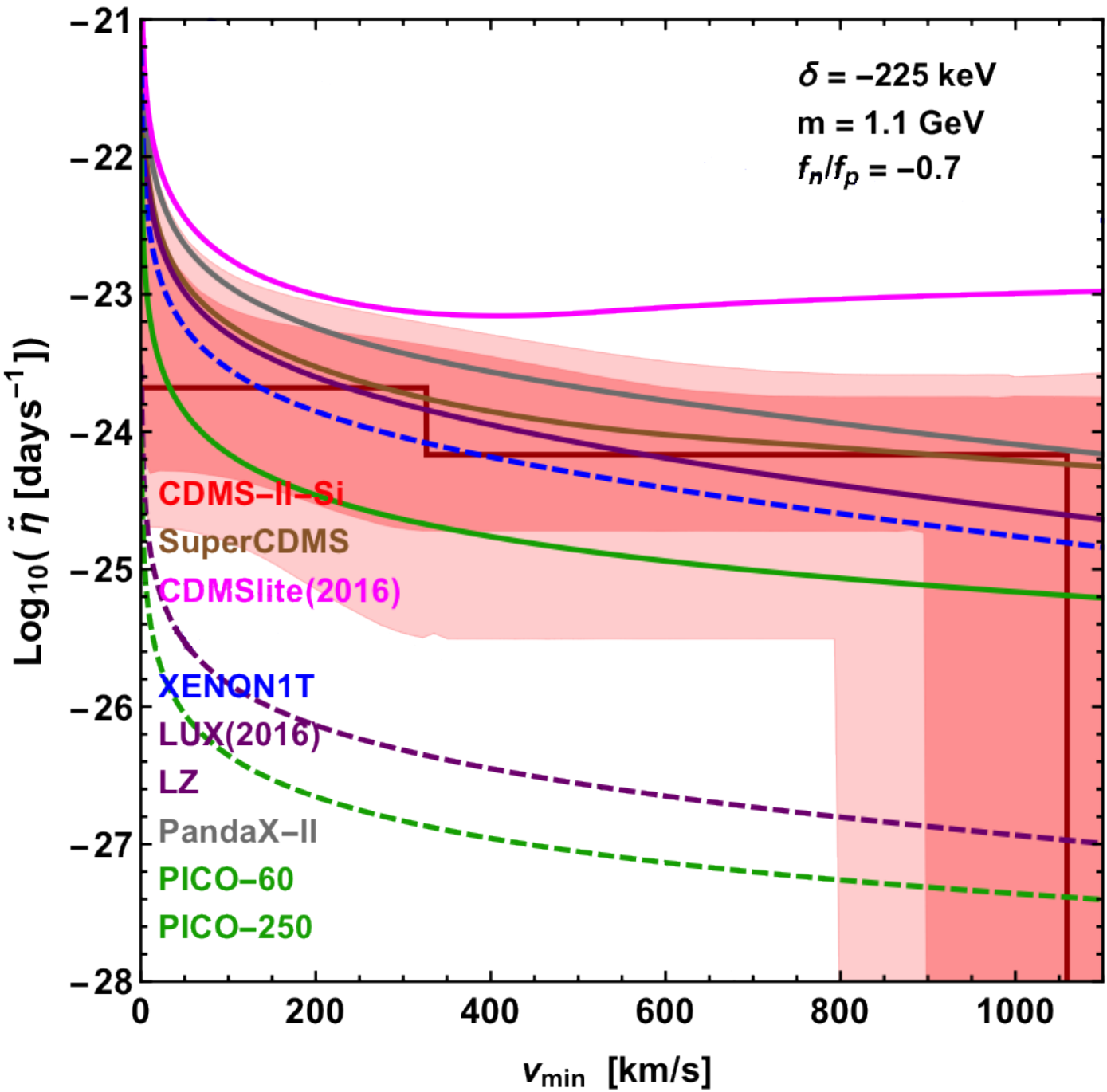}}
\caption{For SI $f_n/f_p$$=-0.7$ interactions and inelastic exothermic scattering with $\delta= -225$ keV, \ref{Fig-3}.a (left) 68\% (dark pink) and 90\%C.L. CDMS-II-Si (light pink)  allowed regions assuming the SHM, and \ref{Fig-3}.b (right panel)  best fit $\tilde\eta^{0}$ function (dark red line) and 68\%C.L. (dark pink) and 90\%C.L. (light pink) confidence bands  in  $v_{\rm min}$, $\tilde\eta^{0}$ space obtained with the EHI method~\cite{Gelmini:2015voa} for the CDMS-II-Si candidates (for $m$= 1.1 GeV). 90\% C.L. present (solid lines) and future sensitivity (dashed lines) limits for the relevant experiments are also shown. 
See~\cite{Witte:2017qsy} for references and details.}
\label{Fig-3}
\end{figure}

\subsection{Light WIMPs which can explain some of the hints }

 The right panels of Figs.~\ref{Fig-1}, \ref{Fig-2} and \ref{Fig-3} show the Halo-Independent data  comparison for the same candidate (assuming a particular light WIMP mass) whose Halo-Dependent analysis assuming  the SHM is shown in the left panels of the same figures: a WIMP with SI  interactions, $f_n/f_p=1$ and elastic scattering  in  Fig.~\ref{Fig-1}~\cite{Gelmini:2014psa}, another also with SI interactions but $f_n/f_p=-0.7$  and exothermic ($\delta <0$) scattering  in  Fig.~\ref{Fig-3}~\cite{Witte:2017qsy},  and a WIMP interacting via an anapole (Anapole DM) and elastic scattering in Fig.~\ref{Fig-2}~\cite{DelNobile:2014eta}.  
 
 The crosses in the right panels of Figs.~\ref{Fig-1} and \ref{Fig-2} represent potential rate and modulation amplitude measurements translated into the $v_{min}, \tilde\eta$ space. The vertical bar of the crosses represent $68\%$ C.L. intervals of the  $\tilde\eta^0$ or  $\tilde\eta^1$ functions in 
 Eq.~(\ref{eta-time}) averaged over the $v_{\rm min}$ range indicated by the horizontal bar.
 
 The pink regions in the left panels and the three red crosses in the right panels in  Figs.~\ref{Fig-1} and~\ref{Fig-2}, corresponding to the three events observed by CDMS-II-Si, are forbidden by upper limits. 
 Even without considering the limits, the Halo-Independent analyses in these figures put in evidence problems in the compatibility of the potential DM signal regions by themselves: the (red) crosses representing the unmodulated rate measurements of CDMS-II-Si are either overlapped or below the  crosses indicating the modulation amplitude data as measured by CoGeNT (blue)  and DAMA (green). Since the annual modulation amplitude of any rate cannot be larger than the average rate itself (the rate cannot be negative), this indicates strong tension between the CDMS-II-Si data on one side, and DAMA and CoGeNT modulation data on the other (these two seem largely compatible). Notice that although for Anapole DM when assuming the SHM (Fig.~\ref{Fig-2}.a) the CDMS-II-Si, CoGeNT and DAMA regions overlap for $m=$ 10 GeV (thus seem compatible) in the Halo-Independent analysis  (Fig.~\ref{Fig-2}.b) the rate measured by CDMS-II-Si seems too low to be compatible with the annual modulation amplitudes observed by DAMA and CoGeNT. We see in these examples that both types of analysis, Halo-Dependent and Halo-Independent,  are complementary in the information they provide.

The candidate  in Fig.~\ref{Fig-3}, where $\delta= -225$ keV,  is still (marginally) viable, i.e. the CDMS-II-Si regions in the left panel and the 90\% C.L. band  in the right panel are not entirely ruled out by present 90\% C.L. bounds (solid lines)~\cite{Witte:2017qsy}. However, the same candidate does not make the DAMA or the CoGeNT signals compatible with other potential signals or with bounds: their regions assuming the SHM are not overlapped with any other and entirely rejected by limits~\cite{Gelmini:2014psa}.  This candidate (with SI interactions, $f_n/f_p=-0.7$  and exothermic scattering) will be entirely rejected as an explanation of the CDMS-II-Si data by a future experiment similar to LZ or PICO-250, whose 90\% CL sensitivity limits (dashed lines) are also shown in  Fig.~\ref{Fig-3} (see  ~\cite{Witte:2017qsy} for details).

\begin{figure}[t]
{\includegraphics[width=0.48\textwidth]{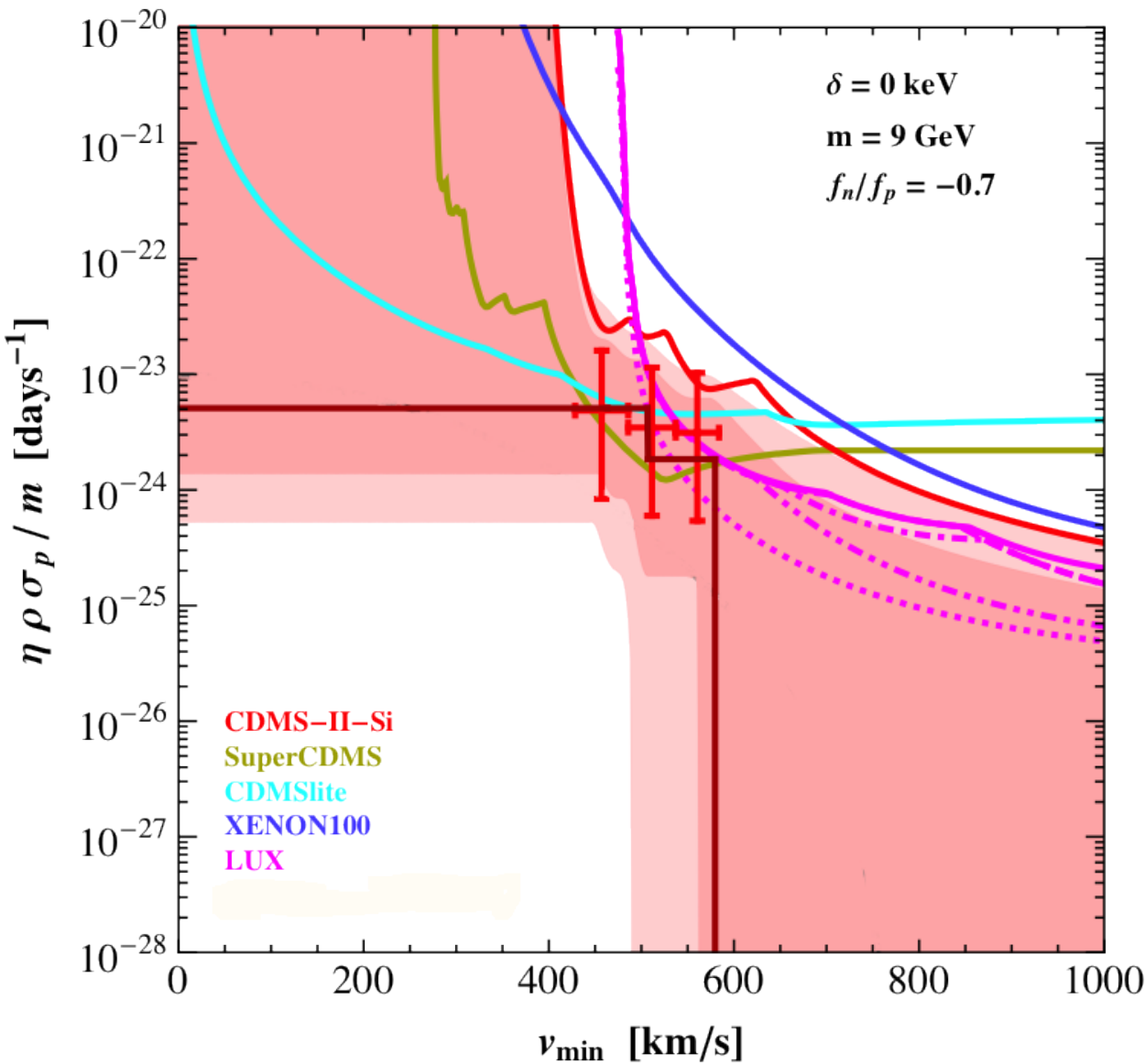}}
{\includegraphics[width=0.46\textwidth]{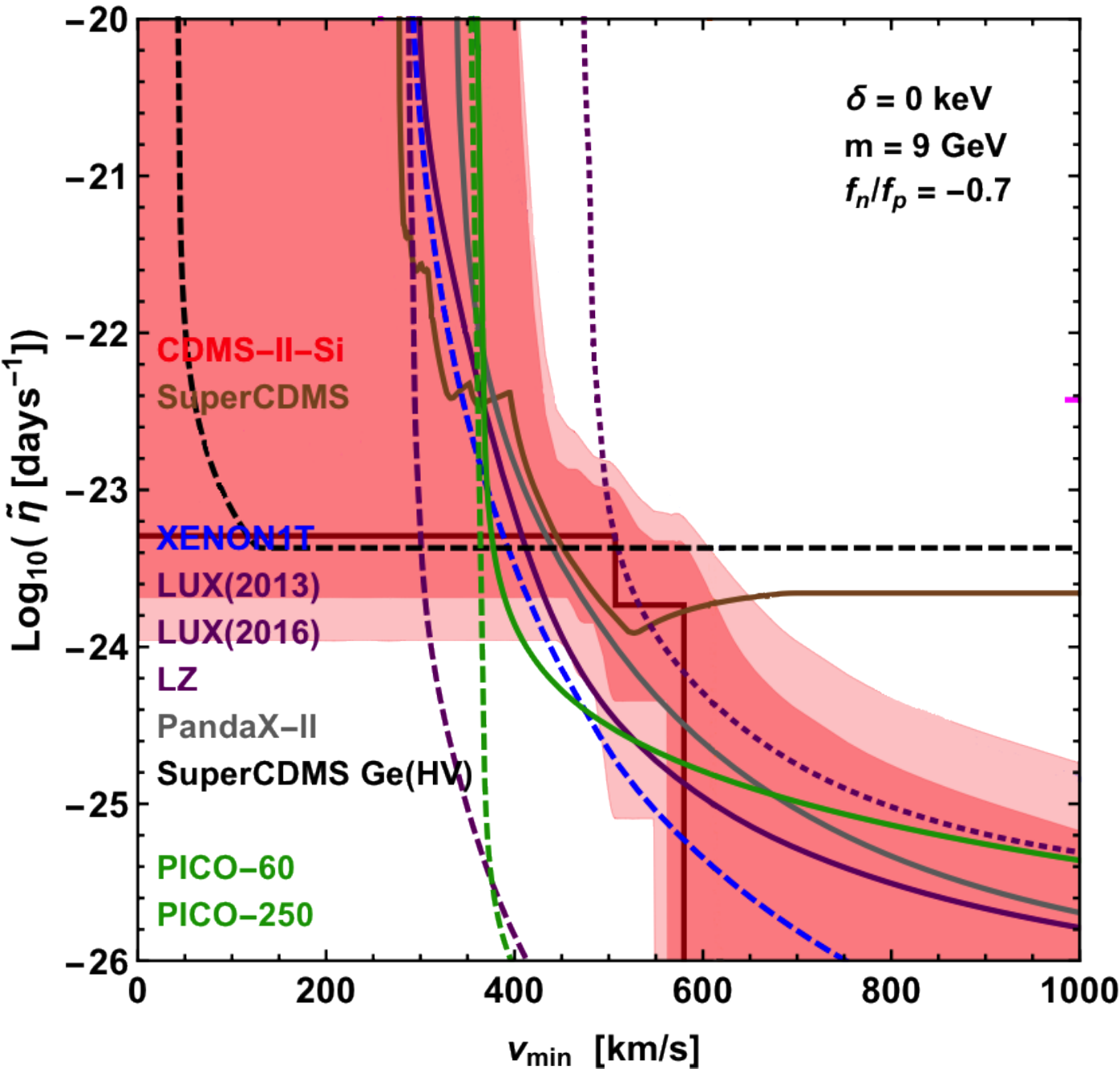}}
\caption{CDMS-II-Si best fit $\tilde\eta^{0}$ function (dark red line) and  68\% and 90\%C.L. (dark  and light pink) confidence bands  in  $v_{\rm min}$, $\tilde\eta^{0}$ space obtained with the EHI method~\cite{Gelmini:2015voa} for a $m=9$ GeV WIMP with elastic  isospin-violating  ($f_n/f_p=-0.7$) SI interaction. \ref{Fig-4}.a (left) shows the best  90\% C.L. limits  as of  July 2015~\cite{Gelmini:2015voa} and \ref{Fig-4}.b (right) shows  90\% C.L present (solid lines) and future sensitivity (dashed lines) limits for the most relevant experiments~\cite{Witte:2017qsy}. Recent LUX and PICO-60 results rejected both bands.}
\label{Fig-4}
\end{figure}

Fig.~\ref{Fig-3}.b shows for  $m=$ 1.1 GeV  the CDMS-II-Si confidence bands of an extended maximum likelihood Halo-Independent analysis (initially proposed in~\cite{Fox:2014kua} and modified in~\cite{Gelmini:2015voa}).   With this method, called EHI (for Extended Halo Independent) in~\cite{Gelmini:2015voa},  a best fit halo function $\tilde\eta^{0}(v_{\rm min})$  (proven to be a piece-wise constant function with downward steps) and a two-sided confidence band are defined. These are shown in Fig.~\ref{Fig-3}.b and Fig.~\ref{Fig-4} as a red line and two pink bands at the 68\% C.L. (darker) and 90\% C.L. (lighter). It is worth noting that the allowed $\tilde\eta^{0}(v_{\rm min})$ functions in the  part of the 90\% C.L. band in Fig.~\ref{Fig-3}.b that escapes all present limits, correspond to halo models different from the SHM~\cite{Witte:2017qsy}. The   $v_{\rm min}$ required here by the highest energy event is improbable in the SHM. 

Fig.~\ref{Fig-4}.a demonstrates that the CDMS-II-Si 68\% C.L. crosses (shown in red in the right panels of  Figs.~\ref{Fig-1} and~\ref{Fig-2}) are similar in vertical extent to the EHI  68\% C.L. confidence band~\cite{Gelmini:2015voa}. In Fig.~\ref{Fig-4} the DM candidate is  an $m=$ 9 GeV light WIMP with SI $f_n/f_p=-0.7$ couplings and elastic interactions.  Both the 68\% and 90\% C.L. bands for this candidate were allowed by the best limits as of July 2015 (shown in Fig.~\ref{Fig-4}.a)  but new LUX 2016 and PICO-60 2017 data  have rejected them, as shown in Fig.~\ref{Fig-4}.b~\cite{Witte:2017qsy}. Notice that a band is rejected when any  $\tilde\eta^{0}$ function  (a continuous function starting at  $v_{\rm min}=0$ and ending with zero value at some maximum $v_{\rm min}$ value) entirely contained in a band must pass above an upper limit.

So far the results of all analyses of compatibility of direct detection data have indicated that only one of the potential direct DM detection signals at a time could be marginally compatible with all negative results for particular DM candidates. For example, a signal in DAMA is favored by a magnetic dipole-moment coupling, or a spin-dependent coupling to protons. Both favor couplings to Na and I and disfavor couplings with Xe and Ge, because the latter have  small magnetic moments and their spin is due mostly to neutrons. These couplings can be combined with inelasticity to further favor a signal in DAMA. Inelastic scattering enhances the potential signal in some targets and suppresses it in some others:  endothermic scattering favors heavier targets (I in DAMA is preferred over Ge in CDMS) while exothermic scattering (as in Figs.~\ref{Fig-3} and \ref{Fig-4}.b)  behaves in the opposite fashion (light nuclei such as Si in CDMS or Na in DAMA are preferred over Xe in LUX or XENON100).  For these reasons, ``Magnetic Inelastic DM (MiDM)"~\cite{Chang:2010en, Barello:2014uda}  or WIMPs with inelastic spin-dependent coupling to protons~\cite{Arina:2014yna, DelNobile:2015lxa} may marginally work as candidates for the DAMA signal. However, there remains a strong tension with null results of other direct detection experiments, specially PICASSO and KIMS, which contain F and I, respectively, whose spin is due mostly to protons. In particular, evading the limits imposed by KIMS (which contains CsI) on the DAMA region requires a different quenching factor for I in both experiments~\cite{Barello:2014uda, DelNobile:2015lxa} (which is not clear experimentally is the case). 

\section{Light WIMPs in Indirect Detection}

There is intense interest at the moment in  an extended GeV $\gamma$-ray excess from the galactic center (GC) detected by FermiLAT which could be a DM signal, in particular a signal of light WIMPs.
Gamma-ray astronomy is done with ground and space instruments. The Fermi Space Telescope was launched in 2008. Its main instrument is the Large Area Telescope (LAT) which detects photons between 20 MeV and 300 GeV. Photons with energy above 20 GeV and up to several TeV are detected by  ground-based Air Cherenkov Telescopes (ACT): HESS in Namibia, MAGIC in Las Palmas and Veritas in the US.  On the planning stage is a large array of ACTs, the CTA (Large Telescope Array), which could detect photons from 10's of GeV to above 100 TeV.
 
 Photons reveal the spatial distribution of their sources because  the Universe is totally transparent to them below 100 GeV. The $\gamma$-ray flux (number per unit area, time and energy)  from  a particular direction expected from DM annihilation is
\begin{equation}
\Phi_{\gamma}(E_{\gamma}) \simeq 
 \left<\sigma_{A} v\right>  \frac{dN_{\gamma}}{dE_{\gamma}} \int_{\rm {line ~of ~sight}} 
 \frac{\rho^2(r)}{m^2} dl(\theta) d\theta ,
 \label{photon-flux}
\end{equation}
where $\left<\sigma_{A} v\right>$ is the  average annihilation cross section times relative speed  at the source, ${dN_{\gamma}}/{dE_{\gamma}}$ is the $\gamma$-ray spectrum per  annihilation (e.g. a delta function for $\chi \chi \to \gamma\gamma$) and the integration of the squared DM number density   as function of distance,  $(\rho(r)/m)^2$,  is along the line of sight and over the angular aperture or resolution of the detector.

Since the annihilation rate depends on the square of the DM density, the rate is boosted in high density regions such as the GC,  DM clumps, dwarf galaxies and other galaxies and galaxy clusters.  Thus a signal is expected preferentially from them.  

Unlike photons, anomalous cosmic rays, such as positrons and antiprotons with energies up to several 100's of GeV  reach us only from short distances within our galaxy. They are an interesting potential signal of WIMP annihilation because there is not much antimatter in the Universe. They are detected from instruments in satellites, such as PAMELA which operated from 2006 to 2011, or AMS-02, mounted on the  International Space Station, in operation since 2011.   Positrons and electrons, $e^+$ and $e^-$, interact with the magnetic fields of the galaxy and rapidly  loose energy within a few kpc. Protons and  antiprotons suffer convective mixing and spallation.  They propagate further than electrons, but still only reach us from a fraction of the size of the galaxy.

\subsection{Light WIMPs in the sky?}

  Subtracting from the Fermi  LAT data all known contributions, D. Hooper with L. Goodenough and later with T. Linden found in 2010~\cite{Goodenough:2009gk} an unexplained extended excess of GeV photons from the GC, peaking at 2-3 GeV. The existence of the excess was  
later confirmed by several other groups (e.g.~\cite{Abazajian:2012pn, Gordon:2013vta, Daylan:2014rsa, Calore:2014xka}) and by the Fermi LAT collaboration itself~\cite{TheFermi-LAT:2015kwa}. 
This signal,  can be interpreted as possible evidence of light WIMPs  with mass close to 10 GeV annihilating predominantly to $\tau^+ \tau^-$ or with mass of  20-45 GeV annihilating predominantly to quarks, with an annihilation cross section  close to  the value of $10^{-26}$ cm$^3$/s required by thermal WIMPs in the early Universe to have the right DM density~\cite{Goodenough:2009gk, Abazajian:2012pn, Gordon:2013vta, Daylan:2014rsa, Calore:2014nla}. 

Many WIMP models could account for this extended excess, but it has been argued that  it  could also be explained by astrophysical sources such as repeated recent outbursts of cosmic-rays from the central 
stellar cluster at the GC~\cite{Carlson:2014cwa, Petrovic:2014uda, Cholis:2015dea} or
 unresolved millisecond pulsars~\cite{Abazajian:2010zy, Bartels:2015aea, Lee:2015fea}. 
 
 The spectral shape of the GC excess is fairly similar to that observed from millisecond pulsars, thus these were proposed earlier on, in 2010, as potential sources of the GeV excess, although several assumptions are needed to fit the observed excess (e.g.~\cite{Abazajian:2014fta, Cholis:2014lta, Hooper:2015jlu, Hooper:2016rap}). Pulsars are rotating neutron stars emitting a beam of electromagnetic radiation.
 Millisecond pulsars have a short rotational period of  1-10 milliseconds. They are believed to be part of binary star systems and to have increased their rotation rates through the accretion of material from their companion stars. Nearly 3000 of them have been detected, but are too faint to be observed as resolved sources in the GC. The electromagnetic spectrum of unresolved millisecond pulsars could explain the extended GeV excess if there would be a large population  of them densely concentrated in a spherically symmetric distribution around the GC. Discussions of the viability of this interpretation center on the existence~\cite{Bartels:2015aea, Lee:2015fea} and the potential origin of such a population of millisecond pulsars~\cite{Cholis:2014lta, Hooper:2015jlu, Hooper:2016rap}. 

\subsection{Indirect detection limits on the GC  GeV $\gamma$-ray excess and light WIMPs}

The GC is a complicated place, with large uncertainties in the DM density profile and many powerful sources.  Other overdense sources in our galaxy, e.g. the inner galaxy and the dwarf satellites of the Milky Way, might provide a cleaner signal.  There is an inner galaxy signal  (within 10 degrees of the GC) consistent with annihilation of  DM particles with the mass and cross section supported by the GC~\cite{Daylan:2014rsa}. However, the GC excess is in tension~\cite{Abazajian:2015raa} with DM annihilation searches from combined dwarf galaxy analyses.

Dwarf galaxies are simpler sources because they are the most DM-dominated structures observed so far. 
The best dwarf  DM limits come from stacked dwarf galaxy  images analyzed by the Fermi-LAT collaboration~\cite{Ackermann:2015zua}.  The tension between the GC and dwarfs data could be alleviated if the halo is extremely concentrated at the GC or the  local DM density is larger than now assumed (so that the DM density at the GC is higher than usually assumed) or the DM annihilation signals from the GC and dwarf galaxies  are somehow different~\cite{Calore:2014nla, Abazajian:2015raa}. The projected sensitivities of 10 y and 15 y of dwarf galaxy observations by FermiLAT~\cite{Charles:2016pgz} will constitute a significant challenge to the DM interpretation of the GC excess.

The ``WMAP Haze" or ``WMAP/Planck Haze" is an excess of microwave emission in the inner 20 degrees (1 kpc)  around the GC, discovered by D. Finkbeiner in 2003~\cite{Finkbeiner:2003im} in WMAP  data, later seen in Planck data too~\cite{:2012rta} (see e.g.~\cite{Egorov:2015eta}).  Planck determined that the spectrum of the Haze corresponds to synchrotron radiation of electrons and positrons accelerated by magnetic fields. The Haze is now considered part of the ``Fermi Bubbles" (discovered in 2009 by Dobler et al.~\cite{Dobler:2009xz, Fermi-LAT:2014sfa}),  two  large structures of $\gamma$-ray emission of  8 kpc in diameter each, on both sides of the galactic plane. Thus, the origin of the emission of the microwaves and gamma rays are supposed to be related, although the origin of both is not known with certainty. They could be  possibly due to an early period of strong jet-like activity of the now dormant  black hole at the GC, or due to repeated starburst events of the central nuclear stellar cluster (see e.g.~\cite{Crocker:2014fla} and \cite{Guo:2016ulg} and references therein). 
In any event, a component of synchrotron radiation from $e^-$ and $e^+$  produced in  DM annihilation (accelerated in magnetic fields) could contribute to the ``Haze" and present upper limits on this component do not contradict the DM parameter regions needed to fit the GC $\gamma$-ray 
excess~\cite{Egorov:2015eta}.

A potential DM hint has also been found in anomalous cosmic rays, not of light WIMPs, but of heavy ones, which if confirmed would imply that light WIMPs are not the dominant form of DM. In this case, light WIMPs could still be detected in direct DM detection experiments (see e.g.~\cite{Duda:2001ae}) but not in indirect detection (due to the $\rho^2$ factor in the rate). An excess in the positron fraction $e^+/(e^+ + e^-)$ in cosmic rays  in the 10 to 100 GeV energy range was reported by PAMELA in 2008~\cite{Adriani:2008zr} (already found in balloon-born experiments since the 1980's) and later confirmed by FermiLAT~\cite{FermiLAT:2011ab} and more recently by  AMS-02~\cite{Aguilar:2013qda}. This is due to an excess of $e^+$ at energies above 10 GeV over what is expected from secondary cosmic rays extending at least to 350 GeV~\cite{Adriani:2013uda, Aguilar:2014mma}, which indicates the existence of nearby primary sources of high energy positrons, such as pulsars or annihilating or decaying DM~\cite{Cholis:2013psa}. No end point to the $e^+$ excess has been observed (by HESS and FermiLAT in the spectrum of  $e^+ + e^-$~\cite{Aharonian:2008aa}) which would be an indication of the mass of the DM particle (this is the maximum energy of any decay or annihilation product).  This requires the DM particle to have a mass above a TeV. Moreover, the sources cannot produce proton-antiproton pairs, because no antiproton excess has been observed, thus the type of DM which could explain the excess is called ``leptophilic". 

If the PAMELA excess is due to DM annihilation, the products should be intermediate light states which then decay into light enough particles, muons or charged pions~\cite{Cholis:2013psa}, and the annihilation cross section should be $\sim 10^{-23}$ cm$^3$/s, larger by several orders of magnitude than expected for thermal WIMPs. For s-wave annihilation ($<\sigma v>$ is $v$ independent), these large cross sections are strongly disfavored by the Planck upper limit~\cite{Ade:2015xua} on DM annihilation at recombination (see below). However other possibilities cannot be excluded, such as 
p-wave annihilation ($<\sigma v> \sim v^2$), because  the speed of DM particles at recombination is many orders of magnitude smaller than in the galactic halo. Besides, constraints imposed by annihilation  in the GC are only compatible with  halo models that predict a relatively small amount of DM in the GC (cored  profiles)~\cite{Meade:2009iu}. If the signal is instead due to DM decay, the required lifetime is $\sim 10^{26}$ s, but limits coming from FermiLAT galaxy clusters and  extragalactic background data practically exclude this lifetime~\cite{Cirelli:2012tf, Ibarra:2013zia} (limits from our galaxy and from CMB measurements are not competitive for decaying DM~\cite{Slatyer:2016qyl}). 

There are DM particle models which can account for all the required properties to explain the PAMELA excess, but  also nearby pulsars remain a potential sources for it (see e.g.~\cite{Cholis:2013psa}). Upcoming AMS-02 data will help to settle the origin of this excess, not only by increasing statistics and extending studies to higher energies, but also by further  constraining any anisotropy in the positron and electron flux. If the origin of the positrons is one of the pulsars nearby, there should be an anisotropy at some level.

Finally let us mention upper limits on the early Universe WIMP annihilation cross section which reject some types of light WIMPs. They  stem from three different types of observations: 
CMB anisotropy precision measurements by several experiments and most recently by the Planck satellite~\cite{Ade:2015xua}, FermiLAT observations of  stacked dwarf galaxy  images~\cite{Ackermann:2015zua} and 
  the positron spectrum measured by AMS-02~\cite{Aguilar:2014mma, Ibarra:2013zia} (see e.g.~\cite{Slatyer:2015jla} and references therein). CMB measurements constrain DM annihilations (or decays) at recombination or after, i.e.  280,000 years after the Bang or later, when $T \leq $ eV. This is the time when injection of secondary particles due to DM annihilation (or decay) affects the process of recombination or heats up the intergalactic medium through which the CMB travels at later times. The dwarf galaxies and positron spectrum observations apply to DM annihilation at present of DM particles bound to galactic haloes.
 
  These upper limits impose $<\sigma v>  < 3 \times10^{-26}$ cm$^3$/s for WIMP masses $m < O(10)$ GeV (the exact mass limit depends on the annihilation mode, see e.g.~\cite{Slatyer:2015jla}). Let us recall that thermal  WIMPs require that  $<\sigma v> \simeq10^{-26}$ cm$^3$/s  at freeze-out  to have the right DM density. With a smaller cross section the relic density predicted exceeds the observed DM density, thus the WIMP candidate is rejected. Thus,  thermal light WIMPs with $m < O(10)$ GeV are rejected,  but only if $<\sigma v>$ is independent of $v$, i.e. for s-wave annihilation. For p-wave annihilation,   $<\sigma v> \sim v^2$ and the limits are not constraining at freeze-out  (see e.g.~\cite{Diamanti:2013bia}), because the WIMP speed $v$  is much larger at freeze-out than at later times. Recall that freeze-out of  thermal WIMPs happens at $T \simeq m/20$,  less than a minute after the Bang. 
   At freeze-out  $v \simeq c/3$ and  then $v$ decreases  as $v \sim T$ due to the adiabatic expansion of the Universe so that it is $v \simeq$ c (10 eV/$m$) at recombination.  WIMPs heat up when they become bound to a halo. Their virial speed  $v$ increases by several orders of magnitude to $v \simeq 10^{-3} c$ in the halo of our galaxy, but still this is much smaller than $v$ at freeze-out.

  P-wave annihilation of thermal WIMPs is one of the many scenarios in which light WIMPs are not rejected  by Planck, FermiLAT or positron spectrum limits on the early Universe WIMP annihilation cross section. WIMPs can be produced in many other non-thermal ways. Most of the light WIMP models explaining the GC GeV~\cite{Calore:2014nla} are consistent with all constraints on DM annihilation in the early Universe~\cite{Ade:2015xua}.
  
\section{Light WIMPs at the LHC}
 
DM particles escape  detection at colliders, thus they are characterized by missing transverse energy (MET) in collider events.  MET detection is difficult  because it requires to measure accurately the energy/momentum of everything visible. Besides, neutrinos also escape from the detectors and are a background to a DM signal, if they cannot be identified by their associated particles.
	
Searches at the LHC can test either complete theories, in which the visible particles which accompany the production of the DM particles are known (such as supersymmetric theories) or, on the opposite side in terms of completeness of the model,  single effective couplings, or, in between both these extremes,  simplified DM models.  The approach of using  a single effective coupling has been used extensively since 2010 to search for generic DM particles at the LHC. This approach is limited because it neglects possible interference between different operators which usually occur in realistic models and also because it assumes only very heavy mediators. Simplified DM models are used to address these limitations.

Limits on light WIMPs have been obtained at the LHC searching for the production of a pair of WIMPs  and one visible particle, emitted either by the initial or the intermediate Standard Model particles, necessary to be able to identify the events. These are called ``mono-X" events~\cite{Beltran:2010ww, Goodman:2010ku, Fox:2012ee}. If the one observable particle is a photon, they are ``monophoton" events; if it is a gluon, they are ``monojet" events. Mono-W's (leptons), mono-Z's (dileptons), or mono-Higgses events are also studied. 
Most limits obtained from these events have been derived assuming a single effective coupling and contact interactions, i.e. heavy mediators of mass $M$ whose propagators reduce to a $1/M^2$  factor. For example,  in the limit $M^2>> q^2$ a factor $g_{SM} g_{DM}/ |q^2 - M^2|$ becomes an effective coupling $1/{M_*}^2$  (here $g_{SM}$ and $ g_{DM}$ are the mediator coupling constants to Standard Model and DM particles, respectively, and $q$ is the momentum transfer).  In this limit, for example  the effective vector coupling of a DM  fermion particle $\chi$ and a quark $\psi_q$ becomes
\begin{equation}
\mathcal{L}_{\rm Effective} = \frac {1}{M_*^{2}}~\bar{\chi} \gamma^\mu \chi ~\bar{\psi_q} \gamma_\mu \psi_q.
 \label{effective coupling}
\end{equation}

The same single effective coupling used to obtain LHC bounds can be used to compute the WIMP interaction with nuclei in direct searches, so the LHC and direct detection limits can be compared. Some plots of this type
(as in~\cite{Fox:2012ee}) show LHC bounds on light WIMPs much more restrictive than direct detection bounds. However, this type of plots must be understood with care. If the local density of the WIMP in question differs by a factor {\it{r}} with respect to the assumed local density, the direct detection  cross section regions and limits must be multiplied by  $1/${\it r}, while the LHC limits do not change. More importantly, while it is valid to include the direct detection limits when presenting the LHC limits derived from contact interactions, the reverse is not correct. The reason is that mediators that are heavy with respect to typical LHC partonic momentum transfers  of O(100 GeV) are also heavy in comparison with the typical O(MeV) momentum transfer in direct detection, but  the opposite is not true and if the mediator is light enough
  the analysis of the collider data is very different.  
 
    For example, putting back the whole propagator $g_{SM} g_{DM}/ |q^2 - M^2|$ in the coupling in Eq.~(\ref{effective coupling}) and varying the vector boson mass $M$, it was shown~\cite{Buchmueller:2013dya} that the validity of the effective coupling in Eq.~(\ref{effective coupling}) in LHC  DM monojet  searches  holds only for $M > 2.5$ TeV. For an  intermediate mass range of lighter mediators (e.g. 0.5 TeV $< M <$ 2.5 TeV for $m=$ 250 GeV), the  effective coupling underestimates  the true cross section upper limit  because the process is resonantly enhanced~\cite{Buchmueller:2013dya}. For yet lighter mediators (e.g.  $M <$ 0.5 TeV for $m=$ 250 GeV), the  effective coupling severely overestimates the true cross section upper limit due to monojet events~\cite{Buchmueller:2013dya}. Combining results from a variety of colliders searches, including  the production of the intermediate vector boson too, it was shown (see e.g.~\cite{Chala:2015ama}) that the parameter space of  simplified models with a vector boson mediator can be much more  tightly constrained. A complete study of the mediator itself in all experiments is not a simple task because  the couplings of the particular mediator with  Standard Model particles must be specified, and  each  effective contact interaction corresponds to many different possible particle models for the mediator. 

   As an example  of  the many possible variations of  a simplified DM model compatible with all known limits on new physics corresponding to a particular contact interaction,   let us consider the simplest model leading to the effective coupling in Eq.~(\ref{effective coupling})  (for more details see e.g.~\cite{Abdallah:2015ter}). It adds to the Standard Model a new gauge vector boson $V$  of a new $U'(1)$ gauge group,  and a DM fermion $\chi$, e.g. a Dirac fermion, assumed to be a singlet under all Standard Model interactions. The DM particle $\chi$  and quarks  carry a charge under the new group, so that both  $\chi$ and the quarks couple to $V$. One needs,  to specify if $V$ couples also to leptons or only to quarks (in which case $V$ is called ``leptophobic") and if its couplings to the three generations of quarks (or quarks and leptons) is the same or not.  Then,  since a mass term for a gauge boson is not renormalizable unless it is generated via spontaneous symmetry breaking, having a mass $M \neq 0$ for $V$ requires adding a Higgs boson field $\Phi$ to break spontaneously the new $U'(1)$. $\Phi$  must have a vacuum expectation value  of the order of magnitude of $M$, $<\Phi> \sim M$. This means that the mass of $\Phi$ cannot be very different from $M$ and, thus, the production of both, $V$ and $\Phi$, needs  to be studied if $M$ is small with respect to typical energies of  colliders. 
      
There are still more necessary choices of this simplified model that lead to different phenomenologies~\cite{Abdallah:2015ter}. 
       Generically, $\Phi$ does not need to couple directly to quarks and leptons, but in principle it mixes with the Standard Model Higgs field and, through this mixing, it couples to all particles in the Standard Model. At this point there are many different possibilities, depending on the assumed type of coupling of $\Phi$ with the Standard Model Higgs. Another choice relates to the origin of the mass $m$ of the DM particle $\chi$. If $\chi$ is a  Dirac fermion, the Higgs field $\Phi$ is also responsible for generating the DM mass $m$. In this case $m$
       is proportional to  $<\Phi>$ and thus to  $M$,   implying that the DM mass cannot be raised arbitrarily compared to the mediator mass.  Also, the model may or not  be affected by higher order corrections. For example, once quarks are coupled to $V$, loop diagrams induce a mixing of $V$ to the Standard Model neutral vector bosons, the photon and the $Z$, and conflicts of these mixings with electroweak precision observables must be taken into account. Another complication arising at one loop is that, depending on the charges of quark and leptons under the new dark $U'(1)$ (e.g. if $V$ is leptophobic), the cancellation of anomalies, which are essential for the renormalizability of the theory, may require the addition of new fermions charged under the $U'(1)$ and the Standard Model gauge group. The masses of these additional fermions are expected to be roughly of the order of $M$ too. The values of the masses of all these new particles are important because if  $V$ is light enough to be produced at colliders,  the resulting phenomenology depends crucially on its decay pattern (same for $\Phi$).  We can see in this one example how many different simplified models need to be studied when moving out of a single effective contact coupling to include lighter mediators.
   
  An enormous amount of work remains to be done in the exploration of simplified DM models at the LHC combined with a other accelerator limits (see e.g. \cite{Abdallah:2015ter,  Bauer:2016gys}).  The difficulties in comparing LHC limits with direct and indirect DM limits has led to  recommendations by the ``LHC Dark Matter Working Group" on how to present  LHC DM limits clarifying the assumptions made in each case~\cite{Boveia:2016mrp}.

\section{Summary and Outlook}

Potential signals for light WIMPs with mass in the GeV to few 10's of GeV range have appeared in several direct DM detection experiments, arousing considerable interest in our field. However, so far not two of them have been found to be compatible with each other and with the upper limits set by the direct searches with negative results,  for all the many possible WIMP candidates studied so far, using either Halo Dependent or Halo Independent analyses. 

The interest in light WIMPs was fueled since 2010 by the unprecedented possible coincidence of four separate experimental hints into tantalizingly close regions of mass and scattering cross section. Before 2010 only DAMA had claimed a potential DM signal for many years, which could have been explained by either light  WIMPs or heavier WIMPs with mass $\sim$100 GeV.

Of the four direct detection experiments that in recent years claimed to have a potential light WIMP signal, one, CRESST-II, with an upgraded detector did not confirm in 2014-2015 the existence of an excess of events over background  attributable to DM they had found in 2010-2011, and the significance of the WIMP signal of another, CoGeNT, decreased with time to end up with  a preference for light WIMP recoils over only background much below the 2$\sigma$C.L. The data of CDMS-II-Si  and DAMA remain significant and WIMP candidates have been studied which can account (marginally for DAMA) for the signal in one of them (but not in both) and escape all the upper bounds from other direct detection experiments. The potential signal of CDMS-II-Si consists of only 3 events obtained with a small exposure of 140.2 kg-days, with a 5.4\% probability of being due to known backgrounds. The DAMA measurement (including the earlier DAMA/NaI and the later DAMA/LIBRA experiments) of an annual modulation of their rate at the 9.3$\sigma$C.L. with the period and phase expected if due to DM   remains baffling for all of us.  DAMA has accumulated an impressive exposure of 1.33 ton year and clearly sees an annual modulation, the question is if it is due to DM. There have been many objections to the DAMA result over the years, many suggesting that the modulation was due to backgrounds, none proven to be correct (see e.g.~\cite{Klinger:2016tlo} and references therein).    

  The situation of light WIMPs in direct detection will be definitely clarified with more data. In particular the low energy threshold direct detection experiments such as CDMSlite at SNOLAB (called SuperCDMS HV)  to start in 2019 (with threshold $<100$ eVee), and PICO, now running with 60 liters and expected to upgrade to up to 500 liters after 2018 (with threshold of O(keV))~\cite{Cushman:2013zza} and the new lower threshold DAMA/LIBRA  will be very important for light WIMPs.  Several years ago DAMA/LIBRA changed its instrumentation to lower its threshold from 2 keVee to 1 keVee~\cite{Bernabei:2016bkl}. Their first lower threshold results are expected in 2017. But the mystery of the DAMA annual modulation will not be entirely resolved unless a new experiment is carried out using the same detector material  DAMA uses,  NaI.  This is the aim of a new collaboration of four experiments (see e.g.~\cite{Castelvecchi:2016}): KIMS, DM-Ice (both in the YangYang Underground Laboratory, South Korea), ANAIS (in the Canfranc Underground Laboratory, Spain) and SABRE (with two sites, the Gran Sasso Laboratory, Italy, and the Stawell Underground Physics Laboratory, Australia). Together these experiments will have about the same amount of NaI that is in DAMA/LIBRA, although, except for SABRE (still in construction) their sensitivity will be inferior to that of DAMA, but conclusive results from them will not be reached in many years.
  
  Significant advances are expected in direct detection in the next decade for all WIMP masses from under GeV to  100 TeV~\cite{Cushman:2013zza}.  We do not know where to expect the WIMP mass and cross section to be, so direct detection experiments will continue  in the next decades until reaching the ``neutrino floor",  the level of cross section at which the signal from background neutrinos (from the Sun,  astrophysical sources and interaction of cosmic rays in Earth's atmosphere) will be much larger than a DM signal.
  
Also in 2010 a potential indirect detection signal in gamma rays of the annihilation of light WIMPs at the galactic center was found in FermiLAT data.  The existence of an extended GeV $\gamma$-ray excess from the galactic center  has been confirmed (by several groups, including the FermiLAT collaboration itself) but its DM origin has not been proven by finding a corresponding signal from other high DM density sources, in particular dwarf galaxies.  It is unclear if new FermiLAT data will conclusively confirm or reject the DM origin of the GeV excess, because of the large astrophysical uncertainties involved, but the projected sensitivities of 10 y and 15 y of dwarf galaxy observations by FermiLAT will constitute the most significant challenge to the DM interpretation of this excess. There are also potential yet undetected astrophysical sources of the excess. 

 It is worth noting that even if light WIMPs would constitute (all or part of) the DM in our galaxy and be detected in direct DM searches, it is not guaranteed that they would be detected in indirect searches. The expected DM signal in direct detection is due to the scattering of WIMPs off protons and/or neutrons, while the signal in indirect detection is due do DM annihilation or decay.  DM particles may neither annihilate nor decay at present, e.g. if the DM consists only of stable particles which differ form their antiparticles. In this regard, direct and indirect searches are independent of each other.

While a signal in indirect searches is not guaranteed by a signal in direct searches, since WIMPs need to interact with nucleons  (thus with its constituents, quarks) to produce a signal in direct detection, any DM which could be observed in direct detection, should  be produced at the LHC in proton-proton collisions. LHC limits using ``mono-X" events and single effective couplings are very constraining on light WIMPs, however they assume that the mediator masses are above several TeV (and neglect interference terms between different operators). For lighter mediators the limits can be much less restrictive, but then many different possibilities for simplified models involving the mediators themselves need to be studied and the work to examine them has only started. The study of simplified DM models at the LHC will continue to shed  light on light WIMP candidates, as the LHC continues producing data in the next decade.  Even if a potential DM particle candidate is identified at the LHC, it will still need to be found where the DM is, in the dark halos of our galaxy, other galaxies and galaxy clusters, to be certain that it constitutes the DM.  In this way searches at colliders,  direct detection and  indirect detection are independent and complementary. 

The hints of light WIMPs in direct and indirect detection just mentioned were the first to be found in several independent experiments pointing to roughly the same DM candidate. This is how signals leading to the discovery of the DM constituent we hope will appear at some point. Even if DM hints did not so far result in a discovery, they have had an enormous impact in our research field. They have sparked a resurgence of new ideas for DM candidates, as well as better ways  to compare data from different experiments and to separate a DM signal from backgrounds.
  Many particle models have been proposed in recent years with the purpose of explaining the DM, instead of the more traditional models made to solve theoretical elementary particle consistency problems. Trying to accommodate these DM hints has opened up our imagination to consider a large variety of potential DM candidates and entire DM sectors, as well as new ways  to test  them.

We have here dealt only with WIMP searches, but  much effort is also devoted to search for other potential DM constituents (see e.g.~\cite{pdg-dark-matter} and \cite{Alexander:2016aln}).

DM searches are advancing fast in all fronts,  dedicated DM experiments,  the LHC, astrophysical observations and modeling. Lots of data necessarily lead to many hints. Hopefully at some point several of them will point to the same DM candidate.

\section*{Acknowledgments}
GBG was supported in part by the  US Department of Energy Grant DE-SC0009937.


\section*{References}
\begin{thebibliography}{10}

\bibitem{pdg-dark-matter} 
C. Patrignani {\it et al.} (Particle Data Group), {\it{ ``The Review of Particle Physics (2016) - 26. Dark Matter''}} Chin. Phys. C, 40, 100001 (2016).

\bibitem{Gelmini:2015zpa} 
  G.~B.~Gelmini,
  {\it{``TASI 2014 Lectures: The Hunt for Dark Matter,''}}
  doi: http://dx.doi.org/10.1142/ 9789814678766\_0012
  [arXiv:1502.01320 [hep-ph]];
  G.~Bertone {\it et al.},
  {\it{``Particle Dark Matter: Observations, Models and Searches,''}}  Cambridge UK: Univ. Pr. (2010) 738 p.; doi:10.1017/CBO 9780511770739.
  
\bibitem{pdg-cosmological-parameters}
C. Patrignani {\it et al.} (Particle Data Group), {\it{ ``The Review of Particle Physics (2016) - 25. The Cosmological Parameters''}} Chin. Phys. C, 40, 100001 (2016).

\bibitem{Ade:2015xua} 
  P.~A.~R.~Ade {\it et al.} [Planck Collaboration],
  {\it{``Planck 2015 results. XIII. Cosmological parameters,''}}
  Astron.\ Astrophys.\  {\bf 594}, A13 (2016)
  [arXiv:1502.01589 [astro-ph.CO]].

\bibitem{Carr:1974nx} 
  B.~Carr and S.~Hawking,
  {\it{``Black holes in the early Universe,''}}
  Mon.\ Not.\ Roy.\ Astron.\ Soc.\  {\bf 168}, 399 (1974).

\bibitem{Kusenko:2013saa} 
  A.~Kusenko and L.~J.~Rosenberg,
  {\it{``Working Group Report: Non-WIMP Dark Matter,ÕÕ}}
ÊÊarXiv:1310.8642 [hep-ph].

\bibitem{Green:2014faa} 
  A.~M.~Green,
  {\it{``Primordial Black Holes: sirens of the early Universe,ÕÕ}}
  Fundam.\ Theor.\ Phys.\  {\bf 178}, 129 (2015)
  doi:10.1007/978-3-319-10852-0\_5
  [arXiv:1403.1198 [gr-qc]].
  
\bibitem{Hannestad:2004px} 
  S.~Hannestad,
  {\it{``What is the lowest possible reheating temperature?,''}}
  Phys.\ Rev.\ D {\bf 70}, 043506 (2004)
  [astro-ph/0403291].

   \bibitem{Pospelov:2000bq}
 M.~Pospelov and T.~ter Veldhuis,
 {\it Direct and indirect limits on the electromagnetic form-factors of WIMPs,}
 Phys.\ Lett.\  B {\bf 480} (2000) 181
 [arXiv:hep-ph/0003010].

\bibitem{Sigurdson:2004zp}
 K.~Sigurdson, M.~Doran, A.~Kurylov, R.~R.~Caldwell and M.~Kamionkowski,
 {\it Dark-matter electric and magnetic dipole moments,}
 Phys.\ Rev.\  D {\bf 70} (2004) 083501
 [Erratum-ibid.\  D {\bf 73} (2006) 089903]
 [arXiv:astro-ph/0406355].
 V.~Barger, W.~Y.~Keung and D.~Marfatia,
 {\it Electromagnetic properties of dark matter: Dipole moments and charge form
 factor,}
 Phys.\ Lett.\  B {\bf 696} (2011) 74
 [arXiv:1007.4345 [hep-ph]].
 E.~Masso, S.~Mohanty and S.~Rao,
 {\it Dipolar Dark Matter,}
 Phys.\ Rev.\  D {\bf 80} (2009) 036009
 [arXiv:0906.1979 [hep-ph]].
 K.~Kumar, A.~Menon and T.~M.~P.~Tait,
 {\it Magnetic Fluffy Dark Matter,}
 JHEP {\bf 1202} (2012) 131
 [arXiv:1111.2336 [hep-ph]].
 V.~Barger, W.~Y.~Keung, D.~Marfatia and P.~Y.~Tseng,
 {\it Dipole Moment Dark Matter at the LHC,}
 Phys.\ Lett.\  B {\bf 717} (2012) 219
 [arXiv:1206.0640 [hep-ph]].
  
  \bibitem{Ho:2012bg}
  C.~M.~Ho and R.~J.~Scherrer,
  {\it Anapole Dark Matter,}
  Phys.\ Lett.\ B {\bf 722} (2013) 341
  [arXiv:1211.0503 [hep-ph]].
 A.~L.~Fitzpatrick and K.~M.~Zurek,
 {\it Dark Moments and the DAMA-CoGeNT Puzzle,}
 Phys.\ Rev.\  D {\bf 82} (2010) 075004
 [arXiv:1007.5325 [hep-ph]].
  M.~T.~Frandsen, F.~Kahlhoefer, C.~McCabe, S.~Sarkar and K.~Schmidt-Hoberg,
  {\it The unbearable lightness of being: CDMS versus XENON,}
  JCAP {\bf 1307} (2013) 023
  [arXiv:1304.6066 [hep-ph]].
  M.~I.~Gresham and K.~M.~Zurek,
  {\it Light Dark Matter Anomalies After LUX,}
  Phys.\ Rev.\ D {\bf 89}, no. 1, 016017 (2014)
  [arXiv:1311.2082 [hep-ph]].
  
  
\bibitem{Feldman:2007wj} 
  D.~Feldman, Z.~Liu and P.~Nath,
  {\it{``The Stueckelberg Z-prime Extension with Kinetic Mixing and Milli-Charged Dark Matter From the Hidden Sector,ÕÕ}}
  Phys.\ Rev.\ D {\bf 75}, 115001 (2007)
  [hep-ph/0702123].
  
\bibitem{Alexander:2016aln} 
  J.~Alexander {\it et al.}
  {\it{``Dark Sectors 2016 Workshop Community Report,''}}
  arXiv:1608.08632 [hep-ph].

\bibitem{DDDM} 
  J.~Fan, A.~Katz, L.~Randall and M.~Reece,
 {\it{``Double-Disk Dark Matter,ÕÕ}}
  Phys. Dark Univ. {\bf 2} 139 (2013)
  [arXiv:1303.1521[astro-ph.CO]] and
  {\it{``Dark-Disk Universe,ÕÕ}}
  Phys.\ Rev.\ Lett.\  {\bf 110} 211302 (2013)
  [arXiv:1303.3271 [hep-ph]].

\bibitem{Clowe:2006eq} 
 D.~Clowe {\it et al.} 
  {\it{``A direct empirical proof of the existence of dark matter,''}}
  Astrophys.\ J.\  {\bf 648}, L109 (2006)
  [astro-ph/0608407].

\bibitem{SIDM}
  D.~N.~Spergel and P.~J.~Steinhardt,
  {\it{``Observational evidence for selfinteracting cold dark matter,ÕÕ}}
  Phys.\ Rev.\ Lett.\  {\bf 84}, 3760 (2000)
  [astro-ph/9909386].

\bibitem{Zavala:2012us} 
  J.~Zavala, M.~Vogelsberger and M.~Walker,
  {\it{``Constraining Self-Interacting Dark Matter with the Milky Way's dwarf spheroidals,ÕÕ}}
  Mon.\ Not.\ Roy.\ Astron.\ Soc.\.: Letters {\bf 431}, L20 (2013)
  [arXiv:1211.6426 [astro-ph.CO]].

\bibitem{Alcock:1998fx} 
  C.~Alcock {\it et al.}  [MACHO and EROS Colls.],
  {\it{``EROS and MACHO combined limits on planetary mass dark matter in the galactic halo,ÕÕ}}
  Astrophys.\ J.\  {\bf 499}, L9 (1998)
  [astro-ph/9803082]; 
  P.~Tisserand {\it et al.} [EROS-2 Collaboration],
  {\it{``Limits on the Macho Content of the Galactic Halo from the EROS-2 Survey of the Magellanic Clouds,''}}
  Astron.\ Astrophys.\  {\bf 469}, 387 (2007)
  [astro-ph/0607207];
  L.~Wyrzykowski {\it et al.},
  {\it{``The OGLE View of Microlensing towards the Magellanic Clouds. IV. OGLE-III SMC Data and Final Conclusions on MACHOs,''}}
  Mon.\ Not.\ Roy.\ Astron.\ Soc.\  {\bf 416}, 2949 (2011)
  [arXiv:1106.2925 [astro-ph.GA]].

\bibitem{Griest:2014cqa} 
  K. Griest, A. Cieplak and M. Lehner,
  {\it{``Experimental Limits on Primordial Black Hole Dark Matter from the First 2 yr of Kepler Data,ÕÕ}}
  Astrophys.\ J.\  {\bf 786}, 158 (2014).

 \bibitem{granularity} 
  J.~Yoo, J.~Chaname and A.~Gould,
  {\it{``The end of the MACHO era: limits on halo dark matter from stellar halo wide binaries,ÕÕ}}
  Astrophys.\ J.\  {\bf 601}, 311 (2004)
  [astro-ph/0307437].

\bibitem{Bird:2016dcv} 
  S.~Bird, I.~Cholis, J.~B.~Mu–oz, Y.~Ali-Ha•moud, M.~Kamionkowski, E.~D.~Kovetz, A.~Raccanelli and A.~G.~Riess,
  {\it{``Did LIGO detect dark matter?,ÕÕ}}
ÊÊPhys.\ Rev.\ Lett.\  {\bf 116}, no. 20, 201301 (2016)
ÊÊ
ÊÊ[arXiv:1603.00464 [astro-ph.CO]].

\bibitem{Abbott:2016blz} 
  B.~P.~Abbott {\it et al.} [LIGO Scientific and Virgo Collaborations],
  {\it{``Observation of Gravitational Waves from a Binary Black Hole Merger,ÕÕ}}
ÊÊPhys.\ Rev.\ Lett.\  {\bf 116}, no. 6, 061102 (2016)
Ê
ÊÊ[arXiv:1602.03837 [gr-qc]]. 

\bibitem{Carr:2016drx} 
  B.~Carr, F.~Kuhnel and M.~Sandstad,
  {\it{``Primordial Black Holes as Dark Matter,ÕÕ}}
ÊÊPhys.\ Rev.\ D {\bf 94}, no. 8, 083504 (2016)
Ê
[arXiv:1607.06077 [astro-ph.CO]].

\bibitem{Brandt:2016aco} 
  T.~D.~Brandt,
  {\it{``Constraints on MACHO Dark Matter from Compact Stellar Systems in Ultra-Faint Dwarf Galaxies,''}}
  Astrophys.\ J.\  {\bf 824}, no. 2, L31 (2016)
  doi:10.3847/2041-8205/824/2/L31
  [arXiv:1605.03665 [astro-ph.GA]].
  
\bibitem{Ricotti:2007au} 
  M.~Ricotti, J.~P.~Ostriker and K.~J.~Mack,
  {\it{``Effect of Primordial Black Holes on the Cosmic Microwave Background and Cosmological Parameter Estimates,''}}
  Astrophys.\ J.\  {\bf 680}, 829 (2008)
  [arXiv:0709.0524 [astro-ph]].
  
\bibitem{fuzzy-DM}
  W.~Hu, R.~Barkana and A.~Gruzinov,
  {\it{``Cold and fuzzy dark matter,''}}
  Phys.\ Rev.\ Lett.\  {\bf 85}, 1158 (2000)
  [astro-ph/0003365].
  
\bibitem{Ringwald:2012hr} 
  A.~Ringwald,
  {\it{``Exploring the Role of Axions and Other WISPs in the Dark Universe,''}}
  Phys.\ Dark Univ.\  {\bf 1}, 116 (2012)
  [arXiv:1210.5081 [hep-ph]].

\bibitem{Hewett:2012ns} 
  J.~L.~Hewett {\it et al.},
  {\it{``Fundamental Physics at the Intensity Frontier,''}}
  doi:10.2172/1042577
  [arXiv:1205.2671 [hep-ex]].
  
    
\bibitem{Baer:2014eja} 
  H.~Baer, K.~Y.~Choi, J.~E.~Kim and L.~Roszkowski,
  {\it{``Dark matter production in the early Universe: beyond the thermal WIMP paradigm,''}}
  Phys.\ Rept.\  {\bf 555}, 1 (2015)
  [arXiv:1407.0017 [hep-ph]].

\bibitem{Gelmini:2006pw} 
  G.~B.~Gelmini and P.~Gondolo,
  {\it{``Neutralino with the right cold dark matter abundance in (almost) any supersymmetric model,''}}
  Phys.\ Rev.\ D {\bf 74}, 023510 (2006)
  [hep-ph/0602230].
  
\bibitem{Nussinov:1985xr} 
  S.~Nussinov,
  {\it{``Technocosmology: Could A Technibaryon Excess Provide A 'natural' Missing Mass Candidate?,''}}
  Phys.\ Lett.\  {\bf 165B}, 55 (1985).
  G.~B.~Gelmini, L.~J.~Hall and M.~J.~Lin,
  {\it{``What Is the Cosmion?,''}}
  Nucl.\ Phys.\ B {\bf 281}, 726 (1987).
  
\bibitem{Petraki:2013wwa} 
  K.~Petraki and R.~Volkas,
  {\it{``Review of asymmetric dark matter,''}}
  Int.\ J.\ Mod.\ Phys.\ A{\bf 28} 1330028 (2013)
  [arX iv:1305.4939 [hep-ph]];
  K.~Zurek
  {\it{``Asymmetric Dark Matter: Theories, Signatures, and Constraints,''}}
  Phys.\ Rept.\  {\bf 537} 91 (2014)
  [arXiv:1308.03 38 [hep-ph]].  

\bibitem{Mayet:2016zxu} 
  F.~Mayet {\it et al.},
  {\it{ ``A review of the discovery reach of directional Dark Matter detection,''}}
  Phys.\ Rept.\  {\bf 627}, 1 (2016)
  [arXiv:1602.03781 [astro-ph.CO]].
  
   \bibitem{Jungman:1995df}
  G.~Jungman, M.~Kamionkowski and K.~Griest,
  {\it{``Supersymmetric dark matter,''}}
  Phys.\ Rept.\  {\bf 267}, 195 (1996)
  [arXiv:hep-ph/9506380] and references therein.

\bibitem{IsospinViolating}
  A.~Kurylov and M.~Kamionkowski,
  {\it{``Generalized analysis of weakly interacting massive particle searches,"}}
  Phys.\ Rev.\ D {\bf 69} 063503 (2004)
  [hep-ph/0307185];  
  J.~L.~Feng, J.~Kumar, D.~Marfatia and D.~Sanford,
  {\it{ Isospin-Violating Dark Matter,}}
  Phys.\ Lett.\ B {\bf 703} 124  (2011)
  [arXiv:1102.4331 [hep-ph]].

\bibitem{Gelmini:2014psa} 
  G.~B.~Gelmini, A.~Georgescu and J.~H.~Huh,
{\it{ ``Direct detection of light Ge-phobic'' exothermic dark matter,''}}
  JCAP {\bf 1407}, 028 (2014)
  [arXiv:1404.7484 [hep-ph]].  
  
\bibitem{DelNobile:2011je} 
  E.~Del Nobile, C.~Kouvaris and F.~Sannino,
  {\it{``Interfering Composite Asymmetric Dark Matter for DAMA and CoGeNT,''}}
  Phys.\ Rev.\ D {\bf 84}, 027301 (2011)
  [arXiv:1105.5431 [hep-ph]].
  
\bibitem{Fitzpatrick:2012ix} 
  A.~L.~Fitzpatrick, W.~Haxton, E.~Katz, N.~Lubbers and Y.~Xu,
  {\it{``The Effective Field Theory of Dark Matter Direct Detection,''}}
  JCAP {\bf 1302}, 004 (2013)
  [arXiv:1203.3542 [hep-ph]].
  
\bibitem{Barello:2014uda} 
  G.~Barello, S.~Chang and C.~A.~Newby,
  {\it{``A Model Independent Approach to Inelastic Dark Matter Scattering,''}}
  Phys.\ Rev.\ D {\bf 90}, no. 9, 094027 (2014)
  [arXiv:1409.0536 [hep-ph]].

 \bibitem{Chang:2010en}
 S.~Chang, N.~Weiner and I.~Yavin,
 {\it Magnetic Inelastic Dark Matter,}
 Phys.\ Rev.\  D {\bf 82} (2010) 125011
 [arXiv:1007.4200 [hep-ph]].

\bibitem{DelNobile:2013cva} 
  E.~Del Nobile, G.~Gelmini, P.~Gondolo and J.~H.~Huh,
  {\it{``Generalized Halo Independent Comparison of Direct Dark Matter Detection Data,''}}
  JCAP {\bf 1310}, 048 (2013)
  [arXiv:1306.5273 [hep-ph]].
 
   \bibitem{TuckerSmith:2001hy}
  D.~Tucker-Smith and N.~Weiner,
{\it{``Inelastic dark matter,''}}
  Phys.\ Rev.\ D {\bf 64}, 043502 (2001)
  [hep-ph/0101138];

\bibitem{Graham:2010ca} 
  P.~W.~Graham, R.~Harnik, S.~Rajendran and P.~Saraswat,
 {\it{``Exothermic Dark Matter,''}}
  Phys.\ Rev.\ D {\bf 82}, 063512 (2010)
  [arXiv:1004.0937 [hep-ph]].

\bibitem{Drukier:1986tm} 
  A.~Drukier, K.~Freese and D.~Spergel,
  {\it{``Detecting Cold Dark Matter Candidates,''}}
  Phys.\ Rev.\ D {\bf 33}, 3495 (1986). 

\bibitem{Savage:2008er} 
  C.~Savage, G.~Gelmini, P.~Gondolo and K.~Freese,
  {\it{``Compatibility of DAMA/LIBRA dark matter detection with other searches,''}}
  JCAP {\bf 0904}, 010 (2009)
  [arXiv:0808.3607 [astro-ph]].
   
\bibitem{Bozorgnia:2012eg} 
  N.~Bozorgnia, G.~B.~Gelmini and P.~Gondolo,
  {\it{``Aberration features in directional dark matter detection,''}}
  JCAP {\bf 1208}, 011 (2012)
  [arXiv:1205.2333 [astro-ph.CO]].
  
\bibitem{Lee:2013wza} 
  S.~K.~Lee, M.~Lisanti, A.~H.~G.~Peter and B.~R.~Safdi,
  {\it{``Effect of Gravitational Focusing on Annual Modulation in Dark-Matter Direct-Detection Experiments,''}}
  Phys.\ Rev.\ Lett.\  {\bf 112}, no. 1, 011301 (2014)
  [arXiv:1308.1953 [astro-ph.CO]].
  
\bibitem{Read:2014qva} 
  J.~I.~Read,
  {\it{``The Local Dark Matter Density,''}}
  J.\ Phys.\ G {\bf 41}, 063101 (2014)
  [arXiv:1404.1938 [astro-ph.GA]].
  
\bibitem{Pato:2015dua} 
  M.~Pato, F.~Iocco and G.~Bertone,
  {\it{``Dynamical constraints on the dark matter distribution in the Milky Way,''}}
  JCAP {\bf 1512}, no. 12, 001 (2015)
  [arXiv:1504.06324 [astro-ph.GA]].
  
  \bibitem{Vogelsberger} 
  M.~Vogelsberger {\it et al.}
  {\it{``Phase-space structure in the local dark matter distribution and its signature in direct detection experiments,''}}
  Mon.\ Not.\ Roy.\ Astron.\ Soc.\  {\bf 395}, 797 (2009)
  [arXiv:0812.0362 [astro-ph]].
  
\bibitem{Purcell:2012sh} 
  C.~W.~Purcell, A.~R.~Zentner and M.~Y.~Wang,
  {\it{``Dark Matter Direct Search Rates in Simulations of the Milky Way and Sagittarius Stream,''}}
  JCAP {\bf 1208}, 027 (2012)
  [arXiv:1203.6617 [astro-ph.GA]].
 
\bibitem{Read:2009iv} 
  J.~I.~Read, L.~Mayer, A.~M.~Brooks, F.~Governato and G.~Lake,
  {\it{``A dark matter disc in three cosmological simulations of Milky Way mass galaxies,''}}
  Mon.\ Not.\ Roy.\ Astron.\ Soc.\  {\bf 397}, 44 (2009)
  [arXiv:0902.0009 [astro-ph.GA]].
 
\bibitem{Lisanti:2011as} 
  M.~Lisanti and D.~N.~Spergel,
  {\it{``Dark Matter Debris Flows in the Milky Way,''}}
  Phys.\ Dark Univ.\  {\bf 1}, 155 (2012)
  [arXiv:1105.4166 [astro-ph.CO]];
  M.~Kuhlen, M.~Lisanti and D.~N.~Spergel,
  {\it{``Direct Detection of Dark Matter Debris Flows,''}}
  Phys.\ Rev.\ D {\bf 86}, 063505 (2012)
  [arXiv:1202.0007 [astro-ph.GA]].
  
\bibitem{Ahlen:1987mn} 
  S.~P.~Ahlen, F.~T.~Avignone, R.~L.~Brodzinski, A.~K.~Drukier, G.~Gelmini and D.~N.~Spergel,
  {\it{``Limits on Cold Dark Matter Candidates from an Ultralow Background Germanium Spectrometer,''}}
  Phys.\ Lett.\ B {\bf 195}, 603 (1987);
  P.~F.~Smith and J.~D.~Lewin,
  {\it{``Dark Matter Detection,''}}
  Phys.\ Rept.\  {\bf 187}, 203 (1990).
    
 \bibitem{Fox:2010bz}
 P.~J.~Fox, J.~Liu and N.~Weiner,
{\it{ ``Integrating Out Astrophysical Uncertainties,''}}
 Phys.\ Rev.\  D {\bf 83}, 103514 (2011)
 [arXiv:1011.1915 [hep-ph]].
 
\bibitem{Frandsen:2011gi}
 M.~T.~Frandsen, F.~Kahlhoefer, C.~McCabe, S.~Sarkar and K.~Schmidt-Hoberg,
 {\it{``Resolving astrophysical uncertainties in dark matter direct detection,''}}
 JCAP {\bf 1201}, 024 (2012)
 [arXiv:1111.0292 [hep-ph]].

\bibitem{Gondolo:2012rs}
 P.~Gondolo and G.~Gelmini,
 {\it{``Halo independent comparison of direct dark matter detection data,''}}
 JCAP{\bf 1212}, 015 (2012)
 [arXiv:1202.6359 [hep-ph]].
 
\bibitem{Cushman:2013zza} 
  P.~Cushman {\it et al.},
  {\it{``Snowmass CF1 Summary: WIMP Dark Matter Direct Detection"}}
  arXiv:1310.8327 [hep-ex].

\bibitem{Bernabei:2013xsa} 
  R.~Bernabei {\it et al.} [DAMA Coll.],
  {\it{``Final model independent result of DAMA/LIBRA-phase1,''}}
  Eur.\ Phys.\ J.\ C {\bf 73}, 2648 (2013)
  [arXiv:1308.5109 [astro-ph.GA]].
  
\bibitem{Aalseth:2010vx}
 C.~E.~Aalseth {\it et al.}  [CoGeNT Coll.],
{\it{ ``Results from a Search for Light-Mass Dark Matter with a P-type Point
 Contact Germanium Detector,''}}
 Phys.\ Rev.\ Lett.\  {\bf 106}, 131301 (2011)
 [arXiv:1002.4703 [astro-ph.CO]].
 
\bibitem{Aalseth:2011wp}
 C.~E.~Aalseth {\it et al.},
{\it{ ``Search for an Annual Modulation in a P-type Point Contact Germanium Dark
 Matter Detector,''}}
 Phys.\ Rev.\ Lett.\  {\bf 107}, 141301 (2011)
 [arXiv:1106.0650 [astro-ph.CO]];
 
\bibitem{Aalseth:2014eft} 
 C.~E.~Aalseth {\it et al.}  [CoGeNT Coll.],
  {\it{``Search for An Annual Modulation in Three Years of CoGeNT Dark Matter Detector Data,''}}
  arXiv:1401.3295 [astro-ph.CO];
{\it{``Maximum Likelihood Signal Extraction Method Applied to 3.4 years of CoGeNT Data,''}}
  arXiv:1401.6234 [astro-ph.CO].

\bibitem{Agnese:2013rvf} 
  R.~Agnese {\it et al.}  [CDMS Coll.],
  {\it{``Silicon Detector Dark Matter Results from the Final Exposure of CDMS II,''}}
  Phys.\ Rev.\ Lett.\  {\bf 111}, 251301 (2013)
  [arXiv:1304.4279 [hep-ex]].
  
\bibitem{Angloher:2014myn} 
  G.~Angloher {\it et al.}  [CRESST-II Coll.],
  {\it{``Results on low mass WIMPs using an upgraded CRESST-II detector,''}}
  Eur.\ Phys.\ J.\ C {\bf 74}, no. 12, 3184 (2014)
  [arXiv:1407.3146 [astro-ph.CO]] and
  {\it{``Results on light dark matter particles with a low-threshold CRESST-II detector,''}}
  Eur.\ Phys.\ J.\ C {\bf 76}, no. 1, 25 (2016)
  [arXiv:1509.01515 [astro-ph.CO]].

\bibitem{Angloher:2011uu}
 G.~Angloher {\it et al.} [CRESST-II Coll.],
{\it{``Results from 730 kg days of the CRESST-II Dark Matter Search,''}}
 Eur.\ Phys.\ J.\  C {\bf 72}, 1971 (2012)
 [arXiv:1109.0702 [astro-ph.CO]].

\bibitem{DelNobile:2014sja} 
  E.~Del Nobile, G.~B.~Gelmini, P.~Gondolo and J.~H.~Huh,
  {\it{``Update on the Halo-Independent Comparison of Direct Dark Matter Detection Data,''}}
  TAUP 2013 Proceedings;
  Phys.\ Procedia {\bf 61}, 45 (2015)
  [arXiv:1405.5582 [hep-ph]].

\bibitem{Lee:1977ua} 
  B.~W.~Lee and S.~Weinberg,
  {\it{``Cosmological Lower Bound on Heavy Neutrino Masses,''}}
  Phys.\ Rev.\ Lett.\  {\bf 39}, 165 (1977).
 
\bibitem{Bottino:2003cz} 
  A.~Bottino, F.~Donato, N.~Fornengo and S.~Scopel,
 {\it{``Light neutralinos and WIMP direct searches,''}}
  Phys.\ Rev.\ D {\bf 69}, 037302 (2004)
  [hep-ph/0307303].

\bibitem{Gelmini:2004gm} 
  G.~B.~Gelmini and P.~Gondolo,
  {\it{``DAMA dark matter detection compatible with other searches,''}}
  hep-ph/0405278;
  P.~Gondolo and G.~B.~Gelmini,
  {\it{``Compatibility of DAMA dark matter detection with other searches,''}}
  Phys.\ Rev.\ D {\bf 71}, 123520 (2005)
  [hep-ph/0504010].
  
  \bibitem{DAMA-chan}  R.~Bernabei {\it et al.},
 {\it{``Possible implications of the channeling effect in NaI(Tl) crystals"}} 
   Eur.\ Phys.\ J.\  C {\bf 53}, 205 (2008)
  [arXiv:0710.0288 [astro-ph]]
  
\bibitem{Bozorgnia:2010xy}
  N.~Bozorgnia, G.~B.~Gelmini, P.~Gondolo,
{\it{ ``Channeling in direct dark matter detection I: chann- eling fraction in NaI(Tl) crystals,''}}
  JCAP {\bf 1011}, 019 (2010)
  [arXiv:1006.3110 [astro-ph.CO]].  
    
\bibitem{Collar:2013gu} 
  J.~I.~Collar,
  {\it{``Quenching and channeling of nuclear recoils in NaI(Tl): Implications for dark-matter searches,''}}
  Phys.\ Rev.\ C {\bf 88}, no. 3, 035806 (2013)
  [arXiv:1302.0796 [physics.ins-det]]. 

\bibitem{DelNobile:2014eta} 
  E.~Del Nobile, G.~B.~Gelmini, P.~Gondolo and J.~H.~Huh,
  {\it{``Direct detection of Light Anapole and Magnetic Dipole DM,''}}
  JCAP {\bf 1406}, 002 (2014)
  [arXiv:1401.4508 [hep-ph]].
  
\bibitem{Witte:2017qsy} 
  S.~J.~Witte and G.~B.~Gelmini,
  {\it{``Updated Constraints on the Dark Matter Interpretation of CDMS-II-Si Data,''}}
  arXiv:1703.06892 [hep-ph].
  
\bibitem{Fox:2014kua} 
  P.~J.~Fox, Y.~Kahn and M.~McCullough,
  {\it{``Taking Halo-Independent Dark Matter Methods Out of the Bin,''}}
  JCAP {\bf 1410}, no. 10, 076 (2014)
  [arXiv:1403.6830 [hep-ph]].
 
\bibitem{Gelmini:2015voa} 
  G.~B.~Gelmini, A.~Georgescu, P.~Gondolo and J.~H.~Huh,
  {\it{``Extended Maximum Likelihood Halo-independent Analysis of Dark Matter Direct Detection Data,''}}
  JCAP {\bf 1511} 038 (2015)
  [arXiv:1507.03902 [hep-ph]].
 
\bibitem{Arina:2014yna} 
  C.~Arina, E.~Del Nobile and P.~Panci,
  {\it{``Dark Matter with Pseudoscalar-Mediated Interactions Explains the DAMA Signal and the Galactic Center Excess,''}}
  Phys.\ Rev.\ Lett.\  {\bf 114}, 011301 (2015)
  [arXiv:1406.5542 [hep-ph]].

\bibitem{DelNobile:2015lxa} 
  E.~Del Nobile, G.~B.~Gelmini, A.~Georgescu and J.~H.~Huh,
  {\it{``Reevaluation of spin-dependent WIMP-proton interactions as an explanation of the DAMA data,''}}
  JCAP {\bf 1508}, no. 08, 046 (2015)
  [arXiv:1502.07682 [hep-ph]]. 
 
\bibitem{Goodenough:2009gk} 
  L.~Goodenough and D.~Hooper,
  {\it{``Possible Evidence For Dark Matter Annihilation In The Inner Milky Way From The Fermi Gamma Ray Space Telescope,''}}
  arXiv:0910.2998 [hep-ph];
  D.~Hooper and L.~Goodenough,
  {\it{``Dark Matter Annihilation in The Galactic Center As Seen by the Fermi Gamma Ray Space Telescope,''}}
  Phys.\ Lett.\ B {\bf 697}, 412 (2011)
  [arXiv:1010.2752 [hep-ph]];
  D.~Hooper and T.~Linden
  {\it{``On The Origin Of The Gamma Rays From The Galactic Center,''}}
  Phys.\ Rev.\ D {\bf 84}, 123005 (2011)
  [arXiv: 1110.0006 [astro-ph.HE]];
  D.~Hooper
  {\it{``The Empirical Case For 10 GeV Dark Matter,''}}
  Phys.\ Dark Univ.\  {\bf 1}, 1 (2012)
  [arXiv:1201.1303 [astro-ph.CO]].
  
\bibitem{Abazajian:2012pn} 
  K.~N.~Abazajian and M.~Kaplinghat,
  {\it{``Detection of a Gamma-Ray Source in the Galactic Center Consistent with Extended Emission from Dark Matter Annihilation and Concentrated Astrophysical Emission,''}}
  Phys.\ Rev.\ D {\bf 86}, 083511 (2012)
  [arXiv:1207.6047 [astro-ph.HE]].
  
\bibitem{Gordon:2013vta} 
  C.~Gordon and O.~Macias,
  {\it{``Dark Matter and Pulsar Model Constraints from Galactic Center Fermi-LAT Gamma Ray Observations,''}}
  Phys.\ Rev.\ D {\bf 88}, no. 8, 083521 (2013)
  Erratum: [Phys.\ Rev.\ D {\bf 89}, no. 4, 049901 (2014)]
  [arXiv:1306.5725 [astro-ph.HE]].  

\bibitem{Daylan:2014rsa}
  T.~Daylan, D.~P.~Finkbeiner, D.~Hooper, T.~Linden, S.~Portillo, N.~L.~Rodd and T.~R.~Slatyer,
  {\it{``The Characterization of the Gamma-Ray Signal from the Central Milky Way: 
  A Compelling Case for Annihilating Dark Matter,''}}
  Phys. Dark Univ.{\bf 12},1 (2016)
  [arXiv:1402.6703 [astro-ph.HE]].
 
\bibitem{Calore:2014xka} 
  F.~Calore, I.~Cholis and C.~Weniger,
  {\it{``Background model systematics for the Fermi GeV excess,''}}
  JCAP {\bf 1503}, 038 (2015)
  [arXiv:1409.0042 [astro-ph.CO]].
  
\bibitem{TheFermi-LAT:2015kwa} 
  M.~Ajello {\it et al.} [Fermi-LAT Coll.]
 {\it{``Fermi-LAT Observations of High-Energy $\gamma$-Ray Emission Toward the Galactic Center,''}}
  Astrophys. J. {\bf 819}, no. 1, 44 (2016)
  [arXiv:1511.02938 [astro-ph.HE]].  
  
\bibitem{Calore:2014nla} 
  F.~Calore, I.~Cholis, C.~McCabe and C.~Weniger,
  {\it{``A Tale of Tails: Dark Matter Interpretations of the Fermi GeV Excess in Light of Background Model Systematics,''}}
  Phys.\ Rev.\ D {\bf 91}, no. 6, 063003 (2015)
  [arXiv:1411.4647 [hep-ph]].
  
\bibitem{Carlson:2014cwa} 
  E.~Carlson and S.~Profumo,
  {\it{``Cosmic Ray Protons in the Inner Galaxy and the Galactic Center Gamma-Ray Excess,''}}
  Phys.\ Rev.\ D {\bf 90}, no. 2, 023015 (2014)
  [arXiv:1405.7685 [astro-ph.HE]].

\bibitem{Petrovic:2014uda} 
  J.~Petrovic, P.~D.~Serpico and G.~Zaharijas,
  {\it{``Galactic Center gamma-ray "excess" from an active past of the Galactic Centre?,''}}
  JCAP {\bf 1410}, no. 10, 052 (2014)
  [arXiv:1405.7928 [astro-ph.HE]].

\bibitem{Cholis:2015dea} 
  I.~Cholis, C.~Evoli, F.~Calore, T.~Linden, C.~Weniger and D.~Hooper,
  {\it{``The Galactic Center GeV Excess from a Series of Leptonic Cosmic-Ray Outbursts,''}}
  JCAP {\bf 1512}, no. 12, 005 (2015)
  [arXiv:1506.05119 [astro-ph.HE]].

\bibitem{Abazajian:2010zy} 
  K.~N.~Abazajian,
  {\it{``The Consistency of Fermi-LAT Observations of the Galactic Center with a Millisecond Pulsar Population in the Central Stellar Cluster,''}}
  JCAP {\bf 1103}, 010 (2011)
  [arXiv:1011.4275 [astro-ph.HE]].
  
\bibitem{Bartels:2015aea} 
  R.~Bartels, S.~Krishnamurthy and C.~Weniger,
  {\it{``Strong support for the millisecond pulsar origin of the Galactic center GeV excess,''}}
  Phys.\ Rev.\ Lett.\  {\bf 116}, no. 5, 051102 (2016)
  [arXiv:1506.05104 [astro-ph.HE]].
  
\bibitem{Lee:2015fea} 
  S.~K.~Lee, M.~Lisanti, B.~R.~Safdi, T.~R.~Slatyer and W.~Xue,
  {\it{``Evidence for Unresolved $\gamma$-Ray Point Sources in the Inner Galaxy,''}}
  Phys.\ Rev.\ Lett.\  {\bf 116}, no. 5, 051103 (2016)
  [arXiv:1506.05124 [astro-ph.HE]].
   
\bibitem{Abazajian:2014fta} 
  K.~N.~Abazajian, N.~Canac, S.~Horiuchi and M.~Kaplinghat,
  {\it{``Astrophysical and Dark Matter Interpretations of Extended Gamma-Ray Emission from the Galactic Center,''}}
  Phys.\ Rev.\ D {\bf 90}, 023526 (2014)
  [arXiv:1402.4090 [astro-ph.HE]].
    
\bibitem{Cholis:2014lta} 
  I.~Cholis, D.~Hooper and T.~Linden,
  {\it{``Challenges in Explaining the Galactic Center Gamma-Ray Excess with Millisecond Pulsars,''}}
  JCAP {\bf 1506}, no. 06, 043 (2015)
  [arXiv:1407.5625 [astro-ph.HE]].
  
\bibitem{Hooper:2015jlu}
  D.~Hooper and G.~Mohlabeng,
  {\it{``The Gamma-Ray Luminosity Function of Millisecond Pulsars and Implications for the GeV Excess,''}}
  JCAP {\bf 1603} (2016) no.03,  049
  [arXiv:1512.04966 [astro-ph.HE]]. 
  
  \bibitem{Hooper:2016rap} 
  D.~Hooper and T.~Linden,
  {\it{``The Gamma-Ray Pulsar Population of Globular Clusters: Implications for the GeV Excess,''}}
  JCAP {\bf 1608}, no. 08, 018 (2016)
  [arXiv:1606.09250 [astro-ph.HE]].

\bibitem{Abazajian:2015raa}
  K.~N.~Abazajian and R.~Keeley,
  {\it{``A Bright Gamma-ray Galactic Center Excess and Dark Dwarfs: Strong Tension for Dark Matter Annihilation Despite Milky Way Halo Profile and Diffuse Emission Uncertainties,''}}
 Phys.\ Rev.\ D {\bf 93}, no. 8, 083514 (2016)
  [arXiv:1510.06424 [hep-ph]].
  
\bibitem{Ackermann:2015zua} 
  M.~Ackermann {\it et al.} [Fermi-LAT Coll.],
  {\it{``Searching for Dark Matter Annihilation from Milky Way Dwarf Spheroidal Galaxies with Six Years of Fermi Large Area Telescope Data,''}}
  Phys.\ Rev.\ Lett.\  {\bf 115}, no. 23, 231301 (2015)
  [arXiv:1503.02641 [astro-ph.HE]].
  
\bibitem{Charles:2016pgz} 
  E.~Charles {\it et al.} [Fermi-LAT Coll.],
  {\it{``Sensitivity Projections for Dark Matter Searches with the Fermi Large Area Telescope,''}}
  Phys.\ Rept.\  {\bf 636}, 1 (2016)
  [arXiv:1605.02016 [astro-ph.HE]].
  

\bibitem{Finkbeiner:2003im}
  D.~P.~Finkbeiner,
  {\it{``Microwave ISM Emission Observed by WMAP,''}}
  Astrophys.\ J.\  {\bf 614}, 186 (2004)
  [astro-ph/0311547].
  
  
\bibitem{:2012rta} 
  P.~A.~R.~Ade {\it et al.} [Planck Coll.],
  {\it{``Planck Intermediate Results. IX. Detection of the Galactic haze with Planck,''}}
  Astron.\ Astrophys.\  {\bf 554}, A139 (2013)
  [arXiv:1208.5483 [astro-ph.GA]].
 
\bibitem{Egorov:2015eta} 
  A.~E.~Egorov, J.~M.~Gaskins, E.~Pierpaoli and D.~Pietrobon,
  {\it{``Dark matter implications of the WMAP-Planck Haze,''}}
 JCAP {\bf 1603}, no. 03, 060 (2016)
  [arXiv:1509.05135 [astro-ph.CO]].
  
\bibitem{Dobler:2009xz} 
  G.~Dobler, D.~P.~Finkbeiner, I.~Cholis, T.~R.~Slatyer and N.~Weiner,
  {\it{``The Fermi Haze: A Gamma-Ray Counterpart to the Microwave Haze,''}}
  Astrophys.\ J.\  {\bf 717}, 825 (2010)
  [arXiv:0910.4583 [astro-ph.HE]];
  M.~Su, T.~R.~Slatyer and D.~P.~Finkbeiner,
  {\it{``Giant Gamma-ray Bubbles from Fermi-LAT: AGN Activity or Bipolar Galactic Wind?,''}}
  Astrophys.\ J.\  {\bf 724}, 1044 (2010)
  [arXiv:1005.5480 [astro-ph.HE]].
  
\bibitem{Fermi-LAT:2014sfa} 
  M.~Ackermann {\it et al.} [Fermi-LAT Coll.],
  {\it{``The Spectrum and Morphology of the Fermi Bubbles,''}}
  Astrophys.\ J.\  {\bf 793}, no. 1, 64 (2014)
  [arXiv:1407.7905 [astro-ph.HE]].
  
\bibitem{Crocker:2014fla} 
  R.~M.~Crocker, G.~V.~Bicknell, A.~M.~Taylor and E.~Carretti,
  {\it{``A Unified Model of the Fermi Bubbles, Microwave Haze, and Polarized Radio Lobes: Reverse Shocks in the Galactic CenterÕs Giant Outflows,''}}
  Astrophys.\ J.\  {\bf 808}, no. 2, 107 (2015)
  [arXiv:1412.7510 [astro-ph.HE]].

\bibitem{Guo:2016ulg} 
  F.~Guo,
  {\it{``The AGN Jet Model of the Fermi Bubbles,''}}
  IAU Symp.\  {\bf 322}, 189 (2017)
  doi:10.1017/S1743921316012023
  [arXiv:1609.07705 [astro-ph.HE]].
  
 
\bibitem{Duda:2001ae} 
  G.~Duda, G.~Gelmini and P.~Gondolo,
  {\it{``Detection of a subdominant density component of cold dark matter,''}}
  Phys.\ Lett.\ B {\bf 529}, 187 (2002)
  [hep-ph/0102200].

\bibitem{Adriani:2008zr} 
  O.~Adriani {\it et al.}  [PAMELA Coll.],
  {\it{``An anomalous positron abundance in cosmic rays with energies 1.5 -100 GeV,''}}
  Nature {\bf 458}, 607 (2009)
  [arXiv:0810.4995 [astro-ph]].
  
\bibitem{FermiLAT:2011ab} 
  M.~Ackermann {\it et al.}  [Fermi LAT Coll.],
  {\it{``Measurement of separate cosmic-ray electron and positron spectra with the Fermi Large Area Telescope,''}}
  Phys.\ Rev.\ Lett.\  {\bf 108}, 011103 (2012)
  [arXiv:1109.0521 [astro-ph.HE]].
  
\bibitem{Aguilar:2013qda} 
  M.~Aguilar {\it et al.}  [AMS Coll.],
  {\it{``First Result from the Alpha Magnetic Spectrometer 
  on the International Space Station: Precision Measurement of the Positron
   Fraction in Primary Cosmic Rays of 0.5  - 350 GeV,''}}
  Phys.\ Rev.\ Lett.\  {\bf 110} 141102 (2013);
  L.~Accardo {\it et al.} [AMS Coll.],
  {\it{``High Statistics Measurement of the Positron Fraction in Primary Cosmic Rays of 0.5 - 500 GeV with the Alpha Magnetic Spectrometer on the International Space Station,''}}
  Phys.\ Rev.\ Lett.\  {\bf 113}, 121101 (2014).
  
\bibitem{Adriani:2013uda} 
  O.~Adriani {\it et al.} [PAMELA Coll.],
  {\it{``Cosmic-Ray Positron Energy Spectrum Measured by PAMELA,''}}
  Phys.\ Rev.\ Lett.\  {\bf 111}, 081102 (2013)
  [arXiv:1308.0133 [astro-ph.HE]];
  
\bibitem{Aguilar:2014mma} 
  M.~Aguilar {\it et al.} [AMS Coll.],
  {\it{``Electron and Positron Fluxes in Primary Cosmic Rays Measured with the Alpha Magnetic Spectrometer on the International Space Station,''}}
  Phys.\ Rev.\ Lett.\  {\bf 113}, 121102 (2014).
  
\bibitem{Cholis:2013psa} 
  I.~Cholis and D.~Hooper,
  {\it{``Dark Matter and Pulsar Origins of the Rising Cosmic Ray Positron Fraction in Light of New Data From AMS,''}}
  Phys.\ Rev.\ D {\bf 88}, 023013 (2013)
  [arXiv:1304.1840 [astro-ph.HE]].
    
\bibitem{Aharonian:2008aa} 
  F.~Aharonian {\it et al.}  [HESS Coll.],
  {\it{``The energy spectrum of cosmic-ray electrons at TeV energies,''}}
  Phys.\ Rev.\ Lett.\  {\bf 101} 261104 (2008)
  [arXiv:0811.3894 [astro-ph]];
  A.~A.~Abdo {\it et al.} [Fermi-LAT Coll.],
  {\it{``Measurement of the Cosmic Ray e$^+$ plus e$^-$ spectrum from 20 GeV to 1 TeV with the Fermi Large Area Telescope,''}}
  Phys.\ Rev.\ Lett.\  {\bf 102}, 181101 (2009)
  [arXiv:0905.0025 [astro-ph.HE]].
  
\bibitem{Meade:2009iu} 
  P.~Meade, M.~Papucci, A.~Strumia and T.~Volansky,
  {\it{``Dark Matter Interpretations of the e$^+$- Excesses after FERMI,''}}
  Nucl.\ Phys.\ B {\bf 831}, 178 (2010)
  [arXiv:0905.0480 [hep-ph]].
  
\bibitem{Cirelli:2012tf} 
  M.~Cirelli,
  {\it{``Indirect Searches for Dark Matter: a status review,''}}
  Pramana {\bf 79}, 1021 (2012)
  [arXiv:1202.1454 [hep-ph]].
  
\bibitem{Ibarra:2013zia} 
  A.~Ibarra, A.~S.~Lamperstorfer and J.~Silk,
  {\it{``Dark matter annihilations and decays after the AMS-02 positron measurements,''}}
  Phys.\ Rev.\ D {\bf 89}, no. 6, 063539 (2014)
  [arXiv:1309.2570 [hep-ph]].
      
\bibitem{Slatyer:2016qyl} 
  T.~R.~Slatyer and C.~Wu,
  {\it{``General Constraints on Dark Matter Decay from the Cosmic Microwave Background,''}}
  Phys.\ Rev.\ D {\bf 95}, no. 2, 023010 (2017)
  arXiv:1610.06933 [astro-ph. CO]. 
  
\bibitem{Slatyer:2015jla} 
  T.~R.~Slatyer,
  {\it{``Indirect dark matter signatures in the cosmic dark ages. I. Generalizing the bound on s-wave dark matter annihilation from Planck results,''}}
  Phys.\ Rev.\ D {\bf 93}, no. 2, 023527 (2016)
  [arXiv:1506.03811 [hep-ph]].
  
\bibitem{Diamanti:2013bia} 
  R.~Diamanti, L.~Lopez-Honorez, O.~Mena, S.~Palomares-Ruiz and A.~C.~Vincent,
  {\it{``Constraining Dark Matter Late-Time Energy Injection: Decays and P-Wave Annihilations,''}}
  JCAP {\bf 1402}, 017 (2014)
  [arXiv:1308.2578 [astro-ph.CO]].
  
  
  \bibitem{Beltran:2010ww} 
  M.~Beltran, D.~Hooper, E.~W.~Kolb, Z.~A.~C.~Krusberg and T.~M.~P.~Tait,
  {\it{``Maverick dark matter at colliders,''}}
  JHEP {\bf 1009}, 037 (2010)
  [arXiv:1002.4137 [hep-ph]].
  
\bibitem{Goodman:2010ku} 
  J.~Goodman, M.~Ibe, A.~Rajaraman, W.~Shepherd, T.~M.~P.~Tait and H.~B.~Yu,
  {\it{``Constraints on Dark Matter from Colliders,''}}
  Phys.\ Rev.\ D {\bf 82}, 116010 (2010)
  [arXiv:1008.1783 [hep-ph]].
  
\bibitem{Fox:2012ee} 
  P.~J.~Fox, R.~Harnik, R.~Primulando and C.~T.~Yu,
  {\it{``Taking a Razor to Dark Matter Parameter Space at the LHC,''}}
  Phys.\ Rev.\ D {\bf 86}, 015010 (2012)
  [arXiv:1203.1662 [hep-ph]].
  
\bibitem{Buchmueller:2013dya} 
  O.~Buchmueller, M.~J.~Dolan and C.~McCabe,
  {\it{``Beyond Effective Field Theory for Dark Matter Searches at the LHC,''}}
  JHEP {\bf 1401}, 025 (2014)
  [arXiv:1308.6799 [hep-ph]].

\bibitem{Chala:2015ama} 
  M.~Chala, F.~Kahlhoefer, M.~McCullough, G.~Nardini and K.~Schmidt-Hoberg,
  {\it{``Constraining Dark Sectors with Monojets and Dijets,''}}
  JHEP {\bf 1507}, 089 (2015)
  [arXiv:1503.05916 [hep-ph]].

\bibitem{Abdallah:2015ter} 
  J.~Abdallah {\it et al.},
  {\it{``Simplified Models for Dark Matter Searches at the LHC,''}}
  Phys.\ Dark Univ.\  {\bf 9-10}, 8 (2015)
  [arXiv:1506.03116 [hep-ph]].
  
\bibitem{Bauer:2016gys} 
  M.~Bauer {\it et al.},
  {\it{``Towards the next generation of simplified Dark Matter models,''}}
  Phys.\ Dark Univ.\  {\bf 16}, 49 (2017)
  [arXiv:1607.06680 [hep-ex]].
  
\bibitem{Boveia:2016mrp} 
  A.~Boveia {\it et al.},
  {\it{``Recommendations on presenting LHC searches for missing transverse energy signals using simplified $s$-channel models of dark matter,''}}
  arXiv:1603.04156 [hep-ex].
  
\bibitem{Klinger:2016tlo} 
  J.~Klinger and V.~A.~Kudryavtsev,
  {\it{``Can muon-induced backgrounds explain the DAMA data?,''}}
  J.\ Phys.\ Conf.\ Ser.\  {\bf 718}, no. 4, 042033 (2016) 
 doi:10.1088/1742-6596/718/4/042033  and
  {\it{``Muon-induced neutrons do not explain the DAMA data,''}}
  Phys.\ Rev.\ Lett.\  {\bf 114}, no. 15, 151301 (2015)
  [arXiv:1503.07225 [hep-ph]].
  
\bibitem{Bernabei:2016bkl} 
  R.~Bernabei {\it et al.},
  {\it{``DAMA/LIBRA results and perspectives,''}}
  Bled Workshops Phys.\  {\bf 17}, no. 2, 1 (2016)
  [EPJ Web Conf.\  {\bf 13}, 60500 (2017)]
  [arXiv:1612.01387 [hep-ex]].
   
\bibitem{Castelvecchi:2016} 
  D.~Castelvecchi,
  {\it{``Controversial dark-matter claim faces ultimate test,"}}
 Nature, Volume 532, Issue 7597, pp. 14-15 (2016).

\end{thebibliography}
\end{document}